\documentclass[sigplan,nonacm]{acmart}

\renewcommand\footnotetextcopyrightpermission[1]{}
\usepackage{amsmath,amsopn}
\usepackage{subcaption}
\usepackage{endnotes,microtype,xspace,graphicx,fancyvrb,fvextra,multirow}
\usepackage{booktabs}
\usepackage{array,underscore,relsize}
\usepackage{fancyhdr}
\usepackage{enumitem}
\usepackage[labelfont=bf,font=small,skip=5pt]{caption}
\usepackage{fp}
\usepackage{siunitx}

\usepackage{tikz}
\usetikzlibrary{positioning, arrows.meta, calc, matrix, fit, backgrounds}

\usepackage{balance}

\usepackage{afterpage}

\usepackage{bm}
\usepackage{makecell}

\newcommand{\sys}{\mbox{\textsc{KVGuard}}\xspace}

\newcommand{\PgdAttn}{\mbox{PagedAttention}\xspace}

\AtBeginDocument{}
\newcommand{\cc}[1]{\mbox{\smaller[0.5]\texttt{#1}}}

\fvset{fontsize=\scriptsize,xleftmargin=8pt,numbers=left,numbersep=5pt}

\input{code/fmt}

\definecolor{kvglinebg}{RGB}{223, 238, 165}

\def\Snospace~{\S{}}

\newcommand{\x}{$\times$\xspace}

\if 0

\fi

\newif\ifdraft\drafttrue
\newif\ifnotes\notestrue
\ifdraft\else\notesfalse\fi

\input{glyphtounicode}
\newcolumntype{R}[1]{>{\raggedleft\let\newline\\\arraybackslash\hspace{0pt}}p{#1}}

\newcommand{\squishlist}{
\begin{itemize}[noitemsep,nolistsep]
  \setlength{\itemsep}{-0pt}
}
\newcommand{\squishend}{
  \end{itemize}
}

\usepackage{tikz}

\newcommand*\BC[1]{%
\begin{tikzpicture}[baseline=(C.base)]
\node[draw,circle,fill=black,inner sep=0.2pt](C) {\textcolor{white}{#1}};
\end{tikzpicture}}

\usepackage{xstring}
\newcommand{\PP}[1]{
\vspace{2px}
\noindent{\bf \IfEndWith{#1}{.}{#1}{#1.}}
}

\newcommand{\V}{\checkmark}

\newcommand{\GB}{\,\text{GB}\xspace}

\newcommand{\boxbeg}{%
\vspace{2px}%
\noindent\begin{tabular}{|l|}\hline
\begin{minipage}{\columnwidth}%
\vspace{2px}%
\noindent
}

\newcommand{\boxend}{%
\vspace{2px}%
\end{minipage}\\ \hline
\end{tabular}%
\vspace{-10pt}%
}

\begin{document}

\sloppy
\title{\sys: A KV Cache Safety Checker for LLM Serving Systems}
% \title{\sys: A Fast KV Cache Safety Checker for LLM Serving Systems}
% \title{\sys: A Memory Safety Checker for KV Cache in LLM Serving Systems}

% when 'make draft'
\ifdefined\DRAFT
 \pagestyle{fancyplain}
 \lhead{Rev.~\therev}
 \rhead{\thedate}
 \cfoot{\thepage\ of \pageref{LastPage}}
\fi

% blind: acmart v2.x requires an \author/\affiliation pair per author;
% the 'anonymous' class option hides these at typeset time.
\author{Jonguk Cheong}
\affiliation{%
  \institution{Seoul National University}
  \country{Seoul, Korea}
}
\email{jdal3031@snu.ac.kr}

\author{Myungbin Hwang}
\affiliation{%
  \institution{Seoul National University}
  \country{Seoul, Korea}
}
\email{auddl1020@snu.ac.kr}

\author{Eulgyu Kim}
\affiliation{%
  \institution{Seoul National University}
  \country{Seoul, Korea}
}
\email{eulgyukim@snu.ac.kr}

\author{Byoungyoung Lee}
\affiliation{%
  \institution{Seoul National University}
  \country{Seoul, Korea}
}
\email{byoungyoung@snu.ac.kr}

\begin{abstract}
Modern LLM serving systems manage the KV cache through \PgdAttn,
a custom GPU allocator that partitions a single CUDA-allocated region
into fixed-size physical blocks
and maps each request to those blocks through a CPU-built block table.
This block-based design exposes a previously unstudied class of memory
safety bugs in the KV cache,
including double-free, use-after-free, out-of-bounds access,
uninitialized read, and permission violation,
which silently corrupt outputs or break request isolation without ever
crashing the system.
Existing memory sanitizers cannot detect them,
because they reason about CUDA-level addresses rather than KV cache
blocks, requests, and block tables,
and their redzone- and quarantine-based designs do not fit the scarce
GPU memory budget of LLM serving.

This paper presents \sys,
the first systematic memory-safety abstraction for KV cache management
in LLM serving systems.
\sys maintains a compact shadow block for each physical KV block,
recording per-request permission and per-entry initialization,
and uses this metadata to perform five safety checks across the
\PgdAttn workflow,
spanning CPU allocation, block table construction and translation,
and GPU-side block table lookup and KV entry access.
Each check fires at the boundary that introduced the violation,
enabling early detection during development
rather than reporting downstream output corruption many requests later.
A prototype of \sys on vLLM and SGLang covers all 29 historical real-world
KV cache safety bugs collected from the two systems' patch histories.
\sys further found 8 previously unknown safety violations while reporting no false positives.
It adds 3.0\% and 7.4\% geomean TPOT overhead on Qwen 2.5 14B on vLLM and
SGLang
and at most 0.11\% of the KV cache in memory overhead,
two orders of magnitude below NVIDIA Compute Sanitizer on the same workloads.
\end{abstract}

\maketitle

\section{Introduction}
\label{s:intro}

Modern LLM serving systems treat the KV cache as a first-class GPU
resource,
since throughput is dominated by how efficiently per-request key and
value vectors are allocated, tracked, and
reused~\cite{kvcache,vllm,zhang2025jenga,sheng2023flexgen,lee2024infinigen}.
To manage this resource at scale,
state-of-the-art systems such as vLLM, SGLang, and TensorRT-LLM adopt
PagedAttention~\cite{vllm,sglang,nvidia-tensorrt-llm}.
PagedAttention reserves one large GPU buffer via \cc{cudaMalloc},
partitions it into fixed-size physical KV cache blocks,
and assigns these blocks to requests through a CPU-side allocator and
a CPU-to-GPU block table that GPU kernels consume at runtime.

This block-based design,
however,
exposes LLM serving systems to a class of memory safety bugs that has
not been studied for the KV cache.
The KV cache management layer is structurally a custom allocator built
on top of a single \cc{cudaMalloc} region,
and is therefore vulnerable to the same bug families as heap allocators:
double-free,
use-after-free,
out-of-bounds access,
uninitialized read,
and permission violation.
For example, one recent vLLM bug leaves stale slot-mapping entries
for the padding rows of a CUDA graph,
so the padding rows write into blocks that belong to live requests,
which read the overwritten values back as their own KV vectors and silently
produce corrupted outputs.
\autoref{t:motiv-bug-list} lists similar real-world bugs across
allocation, block table construction, and GPU kernel access,
all of which corrupt outputs, break request isolation, or leak KV
state without ever crashing the system.

Existing memory safety mechanisms cannot detect these bugs.
The two natural candidates are traditional memory corruption detectors
such as ASan~\cite{asan}, SoftBound~\cite{softbound}, and tagged-memory
systems~\cite{mte,cheri,cets},
and GPU memory debugging tools such as NVIDIA Compute
Sanitizer~\cite{nvidia-compute-sanitizer}.
Both reason about addresses and CUDA-level allocations,
not about KV cache blocks, requests, and block tables.
A wrong access that follows a stale or out-of-bounds block table entry
therefore still produces an in-range address and passes the check.
Their redzone- and quarantine-based designs are also poorly matched to LLM
serving,
where GPU memory is scarce and KV cache capacity directly limits batch
size and context length.

This paper presents \sys,
the first systematic memory-safety abstraction for KV cache management
in LLM serving systems.
The key idea of \sys is to maintain a compact shadow block for each
physical KV block,
and to check the serving system's behavior against this metadata
across the entire KV cache management workflow.
Each shadow block records two fields:
a per-request permission field that captures which requests hold the
block and with what access mode,
and a per-entry initialization field that records whether each KV
entry has been written within the block's current lifetime.
Because the serving system never reuses a request index,
the per-request permission also inherently captures the block's
current lifetime.

Using this shadow metadata,
\sys performs five safety checks at distinct points along the KV
cache management workflow:
CPU allocation events,
CPU block table construction,
CPU-to-GPU block table translation,
GPU block table lookup inside a kernel,
and KV cache entry access inside a kernel.
Each check rejects a wrong allocator event, block table mapping, or
kernel access at the first point where the required facts are
available,
pinpointing each violation at its origin rather than at a downstream
symptom such as a corrupted output token.
This design also avoids the redzones, padding, or delayed reuse that
would consume scarce KV cache space.

The shadow block abstraction is not tied to any single serving system.
It relies only on three properties shared by every \PgdAttn-style
system:
a pool of fixed-size physical KV blocks,
a CPU-built block table that maps logical to physical blocks per
request,
and GPU kernels that consume this block table.
Systems such as vLLM, SGLang, and TensorRT-LLM all share these
properties,
so the same shadow block and the same five checks apply directly.
\sys is therefore positioned as a sanitizer in the style of ASan and
NVIDIA Compute Sanitizer,
targeting development-phase use cases such as continuous integration,
pre-release validation of new kernels, and post-incident debugging,
rather than always-on production.

\sys is implemented on vLLM and SGLang,
two popular open-source LLM serving systems.
The evaluation shows that \sys covers all 29 historical real-world
bugs from the vLLM and SGLang patch histories listed in \autoref{t:motiv-bug-list},
and additionally found 8 previously unknown KV cache safety issues
in current vLLM and SGLang execution paths during end-to-end serving.
Across all evaluated workloads,
\sys reports no false positives,
indicating that its checks fire only on genuine safety violations.
\sys adds 3.0\% and 7.4\% geomean TPOT overhead
on Qwen 2.5 14B on vLLM and SGLang with all checks enabled
and at most 0.11\% of the KV cache in memory overhead,
two orders of magnitude below NVIDIA Compute Sanitizer,
which remains widely used during development despite its slowdown.
The contribution of \sys therefore lies not in production-grade
performance,
but in the KV cache safety abstraction and the early detection it
enables,
which prevents KV cache bugs from remaining latent in the
implementation and silently degrading model output across requests.

This paper makes the following contributions:
\begin{itemize}
\item \textbf{Bug class analysis.}
This paper analyzes KV cache memory safety as a new class of memory
safety problem in LLM serving systems,
characterized by 29 real-world bugs from the vLLM and SGLang patch histories.

\item \textbf{Limitation analysis.}
This paper analyzes why traditional memory corruption detectors and
GPU memory debugging tools cannot detect these bugs.
The root cause is that they reason about addresses and CUDA-level
allocations,
not about KV cache abstractions.

\item \textbf{\sys design.}
This paper presents \sys,
the first systematic memory-safety abstraction for KV cache
management.
\sys combines compact shadow blocks with five checks spanning the CPU
allocator,
the CPU and GPU block tables,
and GPU memory accesses,
catching each violation at the boundary that produced it
rather than as downstream output corruption many requests later.

\item \textbf{Implementation.}
This paper implements \sys on vLLM and SGLang,
covers all 29 historical bugs from the two systems' patch histories,
and further uncovers 8 previously unknown safety violations in
current vLLM and SGLang execution paths during end-to-end serving.
\sys reports no false positives
and adds 3.0\% and 7.4\% geomean TPOT overhead
on Qwen 2.5 14B on vLLM and SGLang with all checks enabled
and at most 0.11\% of the KV cache in memory overhead,
two orders of magnitude below NVIDIA Compute Sanitizer.
\end{itemize}

\section{Background: KV Cache Management}
\label{s:back}
\label{ss:back-kv-management}

KV cache management is the key to efficient LLM serving systems.
Since LLM inference is often
memory-bound~\cite{sheng2023flexgen,zhang2025jenga}, each inference request is
associated with a special memory region called a KV cache~\cite{kvcache}.
The KV cache stores the key and value vectors computed by attention
layers for previous tokens, allowing later decoding steps to avoid
redundant computation.
As a result, serving throughput and reliability highly depend on how
the serving system allocates, tracks, and reuses KV cache memory.

\begin{figure}[t]
\centering
\includegraphics[width=1.0\columnwidth]{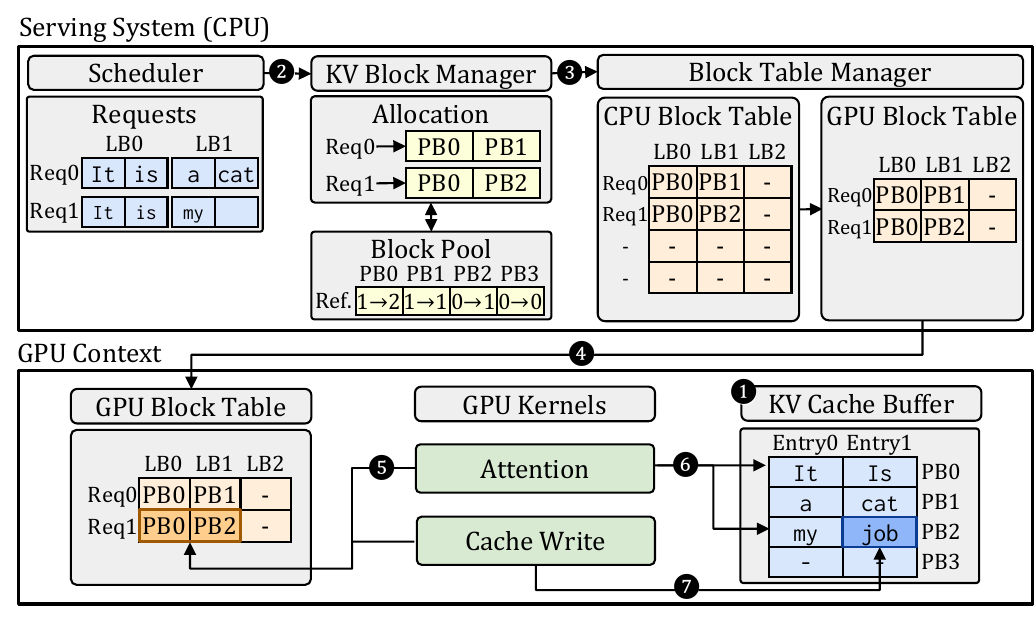}
\caption{Workflow of \PgdAttn.
LB denotes a logical block of a request,
and PB denotes a physical KV cache block in the GPU buffer.
Ref. denotes the reference count of each PB.}
\label{fig:back-pgdattn}
\end{figure}

\PgdAttn~\cite{vllm} is a widely used technique for managing the KV cache.
It introduces a paging mechanism that maps each request's logical blocks
to physical KV blocks in the GPU buffer,
and is widely adopted by state-of-the-art LLM serving systems such as vLLM,
SGLang, and TensorRT-LLM~\cite{vllm,sglang,nvidia-tensorrt-llm}.

\autoref{fig:back-pgdattn} shows the overall workflow of \PgdAttn.
Initially, the serving system allocates a GPU buffer for the KV cache via
\cc{cudaMalloc}~(\BC{1}).
The buffer is divided into KV cache blocks (e.g., from \cc{PB0} to
\cc{PB3}), where each block holds a specific number of KV cache
entries (e.g., two entries in~\autoref{fig:back-pgdattn}).
Serving systems allocate this large buffer once and keep reusing
blocks to avoid repeatedly allocating and freeing GPU memory per
request.

The serving system manages this buffer with a block pool, which records
the reference count of each KV cache block~(\BC{2}).
When a request arrives, the serving system divides the request's input
context into logical blocks, and allocates the required number of KV
cache blocks from the block pool.
For instance, when \cc{Req0} arrives, the serving system allocates two
logical blocks \cc{LB0} and \cc{LB1} for the request, and assigns them
to two physical KV cache blocks \cc{PB0} and \cc{PB1}.
Thus, the block pool is initially updated as \cc{(1, 1, 0, 0)}, meaning
that the first two \cc{PB0} and \cc{PB1} blocks are referenced by
\cc{Req0}.
Then when \cc{Req1} arrives, two logical blocks \cc{LB0} and \cc{LB1}
are allocated, each of which is assigned to \cc{PB0} and \cc{PB2}.
Thus, the block pool is updated to \cc{(2, 1, 1, 0)}, since \cc{PB0} is now
shared by both requests and \cc{PB2} is referenced by \cc{Req1}.

Such request-to-block assignments are written into a CPU block
table~(\BC{3}), which maintains per-request mapping from logical block
indices to physical block indices.
Before launching each kernel, the serving system copies the relevant
CPU block table rows (i.e., the mappings relevant to the requests) to
GPU memory as a GPU block table~(\BC{4}).

After this copy, the GPU access path depends on the GPU block table
rather than the CPU block table.
During attention computation, the GPU attention kernel (denoted as
\cc{Attn.}) reads a GPU block table entry~(\BC{5}), obtains the
physical block index, computes the physical location of the KV cache
entries, and accesses them~(\BC{6}).
Finally, the cache write kernel (denoted as \cc{CW}) writes the newly
computed key and value vectors into the KV cache buffer~(\BC{7}),
using the same GPU-side mapping information to locate the destination
KV cache entry.

\PP{Block Sharing and Permissions}%
Prefix caching~\cite{sglang}, also called prefix sharing, is a mechanism that
allows the serving system to reuse and share KV cache blocks across requests,
demanding read/write permissions for the blocks.
Specifically, when a prefix has already been computed, the same
physical KV block is mapped by logical blocks of multiple requests,
allowing those requests to share the same KV cache entries.
These shared blocks are read-only: later requests may read them during
attention, but they must not overwrite them.
Thus, a request's CPU block table may contain both read-only blocks
(i.e., shared prefix blocks) and allocated blocks that can be written
by that request (i.e., private non-prefix blocks).

\section{Motivation}
\label{s:motiv}

KV cache memory safety matters because LLM serving systems keep per-request
model state for concurrent requests inside a shared GPU allocation.
A wrong KV cache access can corrupt future outputs, or break request
isolation by letting one request read from or write into another
request's KV cache state.
Critically, these failures are hard to notice early because the wrong
KV cache accesses still belong to the allocated KV cache buffer, so
the system may simply produce degraded outputs instead of crashing or
exposing explicit symptoms.
This makes the root cause difficult to identify and fix, and motivates
a memory safety mechanism for the KV cache.
%

% End-to-end bug corpus (2026-10-30). Replaces the former vLLM-only table (fig/motiv/bug-list.tex, no longer input).
% Trailing comments on each row give the corpus ID in ~/research/kvguard and ~/research/kvguard-revival.
% Class rule: a bug with an upstream fix PR (merged or open) by someone other than us is historical (B).
% N rows show the pull request we submitted (all open as of 2026-10-30); N0 shows the
% vLLM developers' fix PR filed in response to our issue #42273 (open).
\begin{table*}[t]
\centering
\caption{Real-world KV cache safety bugs in vLLM and SGLang,
each reproduced end to end and detected by \sys.
% IDs starting with \cc{B} have an upstream fix pull request by other developers.
IDs starting with \cc{N} are newly found by \sys, and Fix PR is the pull request
submitted after the finding.}
\label{t:motiv-bug-list}
\begingroup
\scriptsize
\setlength{\tabcolsep}{1.2pt}
\renewcommand{\arraystretch}{0.88}
\begin{tabular*}{\textwidth}{@{\extracolsep{\fill}}%
>{\raggedright\arraybackslash}p{0.035\textwidth}
>{\raggedright\arraybackslash}p{0.095\textwidth}
>{\raggedright\arraybackslash}p{0.070\textwidth}
>{\raggedright\arraybackslash}p{0.140\textwidth}
>{\centering\arraybackslash}p{0.050\textwidth}
>{\raggedright\arraybackslash}p{0.095\textwidth}
>{\raggedright\arraybackslash}p{0.290\textwidth}
>{\centering\arraybackslash}p{0.080\textwidth}@{}}
\toprule
ID &
System &
Fix PR &
Type &
Check &
\makecell[l]{Model\\Family} &
Workload &
\makecell{Newly\\Found\\by \sys} \\
\midrule
% ---- Historical bugs: vLLM ----
B0 & vLLM & \#47728 & Use-after-free & C1/C5 & Phi-3 & Async scheduling, SWA block pruning & - \\ % VL-H04 (score 4)
B1 & vLLM & \#48085 & Wrong block mapping & C3 & Qwen3.5 & Async scheduling, MTP, CUDA graphs & - \\ % VL-H08 (new impl)
B2 & vLLM & \#36580 & Permission violation & C3/C5 & Llama 3.2 & Chunked prefill, full CUDA graphs & - \\ % VL-H28 (new impl)
B3 & vLLM & \#58259 & Permission violation & C5 & Llama 3.2 & EAGLE3, async scheduling, streaming input & - \\ % VL-D1 (score 5; fixed independently)
B4 & vLLM & \#49146 & Block table OOB & C2 & Phi-3 & CPU KV offload, streaming input & - \\ % VL-D2 (score 5; fixed independently)
B5 & vLLM & \#51900 & Block table OOB & C2/C3 & Llama 3.2 & Streaming input, retained outputs & - \\ % VL-D4 (score 5; independent open PR by KernelClint, 2026-08-12, before our discovery)
B6 & vLLM & \#56709 & Uninitialized read & C5 & Step-3.5 & Multi-module MTP, CPU offload restore & - \\ % F-2026-10-27-04 (new impl; fixed independently)
% ---- Historical bugs: SGLang ----
B7 & SGLang & \#24743 & Permission violation & C3/C5 & Gemma 4 & Hybrid SWA, decode CUDA-graph padding & - \\ % SG-H30 (new impl)
B8 & SGLang & \#1712 & Wrong block mapping & C3 & Qwen2.5 & Overlap scheduling & - \\ % SG-H45 (new impl)
B9 & SGLang & \#2051 & Use-after-free & C1/C3 & Qwen2.5 & Overlap scheduling & - \\ % SG-H48 (new impl)
B10 & SGLang & \#32555 & Use-after-free & C5 & Falcon-H1 & CUDA-graph decode after abort & - \\ % SG-D2 (score 5; fixed independently)
B11 & SGLang & \#40775 & Use-after-free & C1 & Qwen3 & C++ radix tree, shared prefix & - \\ % SG-D4 (score 5; subsystem removed by #40775; cached prefix freed early, its block reallocated to another request while the radix tree still refers to it)
B12 & SGLang & \#39040 & Wrong block mapping & C3 & Mllama & Paged cross-attention decode & - \\ % SG-D5 (score 5; independent issue #38840 and open PR #39040 by SKRohit, 2026-09-10/11)
B13 & SGLang & \#32759 & Uninitialized read & C5 & DeepSeek-V4 & Radix prefix hit, SWA ring & - \\ % SG-D6 (score 4; SWA re-prefill tail restored by amd-danli103, merged 2026-09-07; HIP-only path)
B14 & SGLang & \#38577 & Wrong block mapping & C2 & Llama 3.2 & Multi-LoRA, file-backed HiCache & - \\ % SG-D7 (score 5; fixed independently)
B15 & SGLang & \#45435 & Uninitialized read & C5 & Qwen3.5 & Mixed chunked prefill, overlap scheduling & - \\ % SG-D8 (score 5; fixed independently; re-detected at 0.5.22)
B16 & SGLang & \#32701 & Double free & C1/C2 & Qwen3.5 & Mixed chunks, parallel sampling & - \\ % SG-D9 (score 5; release path dedups the shared page since #32701, merged 2026-07-29; our fork predated it; verified natively 2026-10-30)
B17 & SGLang & \#39283 & Use-after-free & C1 & Qwen3.5 & HiCache load, priority scheduling & - \\ % SG-D10 (score 5; fixed independently; load destination freed while the in-flight host-to-device copy still writes it)
B18 & SGLang & \#36507 & Uninitialized read & C5 & Qwen3.5 & HiCache load, state copy-on-write & - \\ % SG-D11 (score 5; fixed independently)
B19 & SGLang & \#38187 & Wrong block mapping & C4 & Llama 4 & Local attention, FA3, CUDA-graph decode & - \\ % SG-D12 (score 4; fixed independently)
B20 & SGLang & \#42005 & Wrong block mapping & C3 & DeepSeek-V4 & P/D (NIXL), HiSparse & - \\ % SG-D14 (score 5; fixed independently)
B21 & SGLang & \#44081 & Uninitialized read & C5 & Llama 3.2 & EAGLE3, P/D (Mooncake), HiCache load & - \\ % SG-D15 (score 5; fixed independently 2026-10-27)
B22 & SGLang & \#33477 & Uninitialized read & C5 & Falcon-H1 & Overlap, mixed chunk, parallel sampling & - \\ % SG-D16 (score 5; fixed independently)
B23 & SGLang & \#42072 & Out-of-bounds access & C1 & Falcon-H1 & Streaming session close & - \\ % SG-D17 (score 5; fixed independently; sentinel slot -1 freed into the Mamba free list, so a request can be given an out-of-range slot)
B24 & SGLang & \#40223 & Uninitialized read & C5 & DeepSeek-V4 & EAGLE, persistent HiCache hit & - \\ % SG-D20 (score 5; fixed independently 2026-10-12)
B25 & SGLang & \#46990 & Stale block mapping & C3 & Step-3.5 & SWA, EAGLE target verification & - \\ % F-2026-10-28-02 (new impl; fixed independently)
B26 & SGLang & \#46986 & Stale block mapping & C3 & Step-3.5 & SWA, EAGLE draft-extend CUDA graphs & - \\ % F-2026-10-28-03 (new impl; fixed independently)
B27 & SGLang & \#46990 & Stale block mapping & C3 & Gemma 3 & SWA, NGRAM verification & - \\ % F-2026-10-28-04 (new impl; fixed independently)
B28 & SGLang & \#42509 & Uninitialized read & C5 & Llama 3.2 & PDMUX split prefill, EAGLE3 & - \\ % F-2026-10-28-01 (new impl; same guard in open draft PR #42509 by Li-brua, 2026-10-04/08, before our discovery)
\midrule
% ---- New findings: vLLM ----
N0 & vLLM & \#42276 & Stale block mapping & C2 & Phi-3 & Chunked prefill, SWA eviction & \V \\ % paper N0 (our issue #42273; fix PR by a vLLM developer, open)
N1 & vLLM & \#62871 & Wrong block reuse & C1 & Llama 3.2 & Async scheduling, rejected stream chunk & \V \\ % VL-D3 (score 5; our PR, open)
N2 & vLLM & \#64649 & Admission stall & C1 & Qwen3 & Async P2P KV load, speculative decoding & \V \\ % F-2026-10-29-01 (new impl; adapter capacity/progress rule on the allocation path, outside P1-P4: an allocation refused with every block free; our PR, open)
% ---- New findings: SGLang ----
N3 & SGLang & \#45604 & Uninitialized read & C5 & Qwen3.5 & int8 state checkpoint, short request & \V \\ % SG-D1 (score 4; our PR, open; re-detected at 0.5.22)
N4 & SGLang & \#45605 & Permission violation & C3 & Llama 4 & P/D (NIXL), parallel sampling & \V \\ % SG-D13 (score 5; our PR, open)
N5 & SGLang & \#47390 & Use-after-free & C5 & Llama 3.2 & Speculative grammar decoding, overlap scheduling & \V \\ % SG-D18 (score 5 with EAGLE3 on the fork; reproduced natively with NGRAM on 2026-10-30; our PR, open)
N6 & SGLang & \#47392 & Stale block mapping & C2 & Qwen3.5 & NGRAM, streaming session & \V \\ % SG-D19 (score 5; reproduced natively on 2026-10-30, also for stop-ended turns without speculation; our PR, open)
N7 & SGLang & \#47250 & Use-after-free & C1 & Llama 3.2 & P/D, EAGLE3, prefill cancellation & \V \\ % F-2026-10-29-03 (new impl; our PR, open)
% TensorRT-LLM findings F-2026-10-27-01 and F-2026-10-27-03 are excluded from this paper (owner decision, 2026-10-30).
% Withdrawn 2026-10-30: F-2026-10-27-05 (former N3; KVGuard reports were false positives, see
% ~/research/kvguard-revival/results/2026-10-30-vllm-store-reuse-oracle/).
\bottomrule
\end{tabular*}
\endgroup
\end{table*}

\subsection{Case Studies of Real-world KV Cache Safety Bugs}
\label{ss:motiv-bugs}

The KV cache management layer is structurally a custom memory
allocator.
The serving system partitions one \cc{cudaMalloc} region into
software-defined KV cache blocks.
Then it manages those KV cache blocks with allocator primitives, where
its interfaces are similar to \cc{malloc} and \cc{free} but with
different semantics.

Thus, KV cache bugs exhibit a variety of allocator-style bugs, similar
to those in traditional allocators yet with different root causes and
consequences.
Popular open-source LLM serving systems,
such as vLLM~\cite{vllm} and SGLang~\cite{sglang},
have experienced many KV cache safety bugs,
which are summarized in~\autoref{t:motiv-bug-list}.
%
% These includes including  use-after-free,
% out-of-bounds access, uninitialized read, and permission violation, as
% well as stale metadata bugs that make block tables or prefix/offload
% state disagree with actual KV block ownership.
%
To showcase how these bugs violate KV cache safety,
this section presents three case studies.

\afterpage{%
\begin{figure*}[t]
\centering
\includegraphics[width=0.2\textwidth]{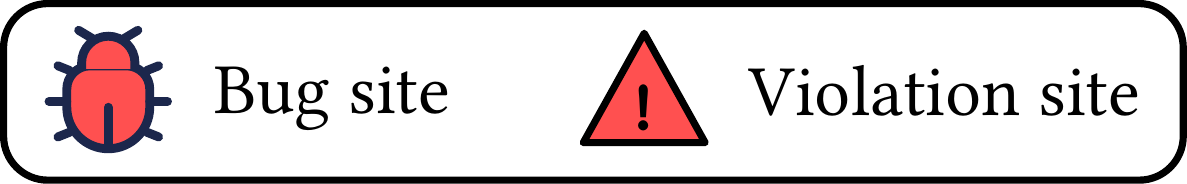}
\par\medskip
\begin{subfigure}[t]{\textwidth}
\centering
\includegraphics[width=\textwidth]{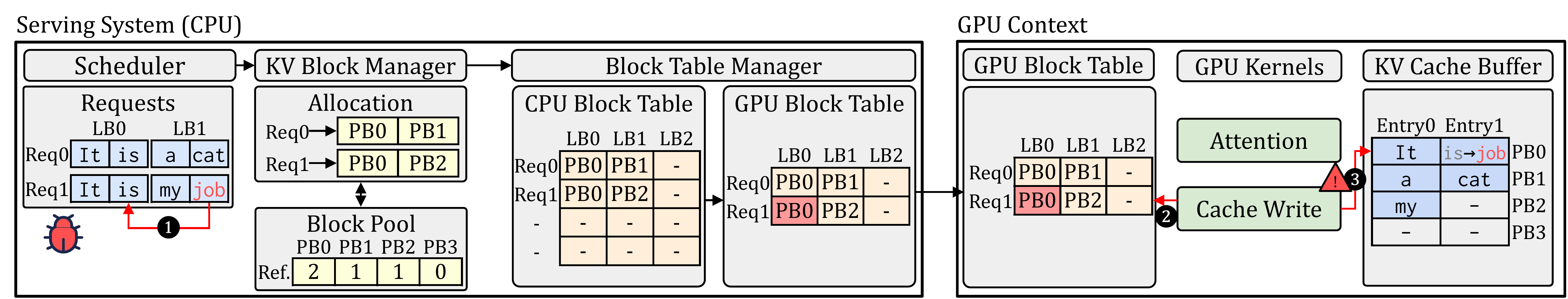}
\caption{Write to a read-only block.}
\label{fig:motiv-case-study-readonly-write}
\end{subfigure}\\[0.0em]
\begin{subfigure}[t]{\textwidth}
\centering
\includegraphics[width=\textwidth]{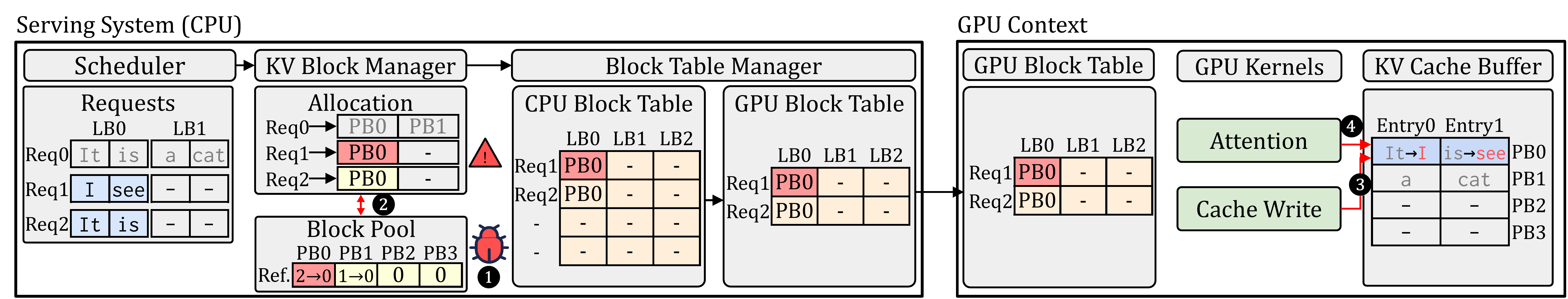}
\caption{Use-after-free of a cached block.}
\label{fig:motiv-case-study-uaf}
\end{subfigure}\\[0.0em]
\begin{subfigure}[t]{\textwidth}
\centering
\includegraphics[width=\textwidth]{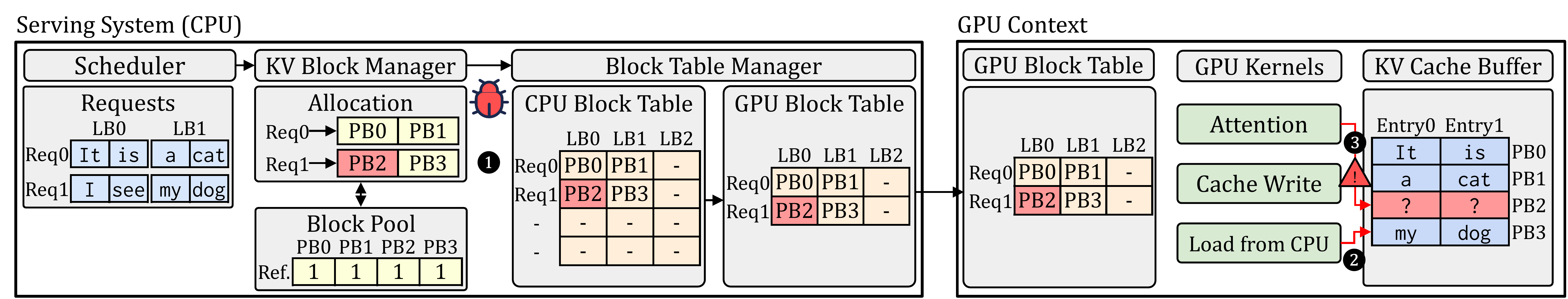}
\caption{Uninitialized read of a block.}
\label{fig:motiv-case-study-uninit-read}
\end{subfigure}
\caption{Three real-world KV cache safety bugs in vLLM and SGLang.
Each case shows a different way that a valid GPU address can still produce
unsafe KV cache behavior.}
\label{fig:motiv-case-studies}
\end{figure*}%
}

\PP{Case 1. Write to a Read-Only Block (B3)}%
\autoref{fig:motiv-case-study-readonly-write} illustrates a bug where
a request writes its new KV vectors into a shared, read-only block.
In the example,
\cc{Req0} and \cc{Req1} share the prefix \cc{It is},
so both requests map their \cc{LB0} to \cc{PB0},
and the reference count of \cc{PB0} is two.
Thus, \cc{PB0} is read-only,
and \cc{Req1} should write its next token \cc{job}
into the second entry of its private block \cc{PB2} (mapped by \cc{LB1}).
However, the bug is that
the scheduler moves the write position of \cc{Req1}
back into a block that is already read-only.
Specifically, the scheduler moves the write position of \cc{Req1} back by two
entries, from \cc{LB1} into \cc{LB0}~(\BC{1}).
Following this position,
the cache write kernel translates \cc{LB0}
through the GPU block table row of \cc{Req1}
and obtains \cc{PB0}~(\BC{2}).
It then writes the KV vectors of \cc{job}
into the second entry of \cc{PB0},
overwriting those of \cc{is}~(\BC{3}).
As a result,
\cc{Req1} violates the read-only permission of \cc{PB0},
and both \cc{Req0} and every later request that reuses the prefix
read \cc{job} in place of \cc{is}.

% \PP{Case 2. Out-of-Bounds Read of a GPU Block Table (B3)}%
% \autoref{fig:motiv-case-study-oob-translation} illustrates a bug where a GPU
% kernel reads outside the GPU block table row for the current request.
% %
% In the example, Req0 has a mapping to \cc{(PB0, PB1)} in its GPU block
% table row.
% %
% Req1 has the one to \cc{(PB2, PB3)}.
% %
% Thus, when the GPU kernel for Req0 runs, it should accordingly
% read Req0's GPU block table row, \cc{PB0} (mapped by \cc{LB0})
% and \cc{PB1} (mapped by \cc{LB1}).
% %
% However, the bug is that Req0's kernel incorrectly checks the row
% length and fails to stop after reading \cc{PB1}.
% %
% Instead, Req0's kernel continues to read the next entry in memory
% (i.e., \cc{PB2}, mapped by \cc{Req1}'s \cc{LB0}), as if it were still
% reading Req0's GPU block table row~(\BC{1}).
% %
% As a result, Req0 reads Req1's KV vectors, which is an out-of-bounds
% KV cache access.

\PP{Case 2. Use-After-Free of a Cached Block (B11)}%
\autoref{fig:motiv-case-study-uaf} illustrates a bug where
the serving system frees a block that the prefix cache still holds.
In the example, \cc{Req0} has finished,
and the serving system keeps the KV vectors of the prefix "\cc{It is}"
in \cc{PB0} so that later requests can reuse them.
Thus, when \cc{Req0} finishes,
the serving system should drop only the reference of \cc{Req0} to \cc{PB0},
decreasing its reference count from two to one,
and only free the private block \cc{PB1}.
However, the bug is that
the serving system misses that \cc{PB0} has been cached for future reuse.
It frees \cc{PB0} as if it were another private block of \cc{Req0},
setting its reference count to zero~(\BC{1}).
When \cc{Req1} arrives,
the block pool hands the freed \cc{PB0} to \cc{Req1},
although the prefix "\cc{It is}" is still mapped to \cc{PB0}~(\BC{2}).
The cache write kernel then writes the KV vectors of "\cc{I see}" for \cc{Req1}
into \cc{PB0},
overwriting the cached KV vectors of "\cc{It is}"~(\BC{3}).
As a result,
any later request that reuses the cached prefix "\cc{It is}"
reads the KV vectors of \cc{Req1}.
For example, when \cc{Req2} arrives with the prefix "\cc{It is}",
it reuses \cc{PB0} due to the prefix match,
and its attention kernel reads the KV vectors of "\cc{I see}"
in place of those of "\cc{It is}"~(\BC{4}),
which is a use-after-free of a KV cache block.
Here, the mapping from "\cc{It is}" to \cc{PB0}
acts like a dangling pointer in a conventional use-after-free bug,
since it still points to \cc{PB0} after the block is freed and reused.
%

% \PP{Case 3. Stale Mapping in a CPU Block Table (B0)}%
% \autoref{fig:motiv-case-study-stale-mapping} illustrates a bug where a reused
% CPU block table row carries stale physical block indices from an old request.
% %
% In the example,
% Req0 previously used \cc{PB0} and \cc{PB1},
% but it is already finished and removed from the serving system's internal
% state.
% %
% When Req1 arrives, the serving system sets up a new CPU block table row
% for Req1.
% %
% The bug is that the serving system incorrectly sets up this new row.
% %
% The row should contain only \cc{PB2}
% because Req1 needs only one block.
% %
% However, due to the bug,
% the serving system appends the stale \cc{PB1}
% from Req0's old row to the new row after \cc{PB2}~(\BC{1}).
% %
% This stale row is then copied to the GPU as a GPU block table for
% Req1~(\BC{2}), and when the GPU kernel for Req1 runs, it reads the
% stale \cc{PB1} entry as if it were still part of Req1's GPU block table
% row~(\BC{3}).
% %
% As a result, Req1 reads (and writes) the stale KV cache block, where
% its lifetime is already expired as Req0 is already finished.

\PP{Case 3. Uninitialized Read of a Block (B6)}%
\autoref{fig:motiv-case-study-uninit-read} illustrates a bug where
a request reads a block that has never been written.
When GPU memory runs short,
serving systems can offload the KV cache blocks of earlier prompts to CPU memory
and load them back into GPU memory
when a later request shares the same prefix.
In the example,
the KV cache blocks of the prompt of \cc{Req1}, "\cc{I see my dog}",
are found in CPU memory.
Thus, the serving system assigns \cc{PB2} and \cc{PB3}
to \cc{LB0} and \cc{LB1} of \cc{Req1},
counts both blocks as computed,
and should load both blocks from CPU memory.
However, the bug is that
the serving system loads only \cc{PB3}
but still counts \cc{PB2} as computed~(\BC{1}).
It decides which blocks to load by the attention window of the model,
and misses \cc{PB2}, an older block that \cc{Req1} still reads.
Thus, the load fills \cc{PB3} with the KV vectors of "\cc{my dog}",
while \cc{PB2} keeps whatever it held before~(\BC{2}).
When the attention kernel later runs for \cc{Req1},
it reads \cc{PB2} as if it held the KV vectors of "\cc{I see}"~(\BC{3}).
As a result,
\cc{Req1} reads uninitialized KV vectors.
Worse, the serving system may later share \cc{PB2} through the prefix cache
as a complete block,
spreading the uninitialized KV vectors to other requests.

% %
% When Req0 finishes,
% the serving system removes Req0 from its internal state and returns Req0's
% blocks to the block pool.
%
% However, the bug leaves Req0's CPU block table row uncleared~(\BC{1}).
% %
% Later, Req1 arrives with a shorter input and needs only \cc{PB2}.
% % %
% The reused row should contain \cc{PB2} followed by empty entries,
% but the stale tail still contains Req0's old \cc{PB1}.
% %
% Thus, Req1's CPU block table row becomes \cc{[PB2, PB1, -]} instead of
% \cc{[PB2, -, -]},
% and this stale row is copied to the GPU as a GPU block table~(\BC{2}).

% If the GPU access path consumes the stale entry,
% Req1 follows \cc{PB1} as if it were Req1's block~(\BC{3}).
%
% As a result,
% Req1 can read or write a KV cache block from Req0's old lifetime,
% or a block that has already been reused by another request.
%

\subsection{Limitations: Using Existing Mechanisms for KV Cache Safety}
\label{ss:motiv-limitations}

A natural question is whether existing memory safety mechanisms can
detect the KV cache bugs listed in~\autoref{t:motiv-bug-list}.
There are two classes of candidates:
traditional memory corruption detectors and GPU memory debugging
tools.
However, neither captures the KV cache abstraction that the case
studies violate.
Both classes reason about addresses and CUDA-level allocations.
They do not reason about KV cache-specific abstractions,
such as KV cache blocks, requests, and block tables.

\PP{Limitations of Traditional Memory Corruption Detectors}%
Traditional detectors such as ASan~\cite{asan}, SoftBound~\cite{softbound}, and
tagged-memory systems~\cite{mte, cets, cheri} enforce memory safety using
per-object or per-pointer metadata.
However, this metadata is attached to addresses, pointers, or
allocator-visible objects.
It is not attached to the software-defined KV cache blocks carved out
of a single \cc{cudaMalloc} region.
Thus, the metadata cannot express whether \cc{Req0} is permitted to
access \cc{PB1}.
It also cannot express whether a GPU block table entry maps the block
that the CPU block table intended.
As a result, the write to a read-only block in Case 1
and the use-after-free in Case 2 appear legal,
because they stay inside one live allocation.
The ASan-style sanitizers are not a good fit either.
LLM serving treats GPU memory as a scarce resource, which directly
impacts the batch size and context length managed by the KV cache.
However, ASan-style sanitizers impose high memory overhead through
redzones and quarantines.

\PP{Limitations of GPU Memory Debugging Tools}%
GPU memory debugging tools such as NVIDIA Compute
Sanitizer~\cite{nvidia-compute-sanitizer} extend these checks to GPU
kernels.
Its \cc{memcheck} tool reports out-of-bounds accesses,
and its \cc{initcheck} tool reports uninitialized reads at the CUDA
allocation level.
However, these tools share the same abstraction mismatch.
They only see CUDA-visible allocations (i.e., an entire KV cache
buffer), not the individual KV cache blocks.
For example, \cc{memcheck} treats the entire KV cache buffer as one
valid region.
Thus, a kernel that follows a stale or out-of-bounds KV block still
appears to be following a valid address, so it passes the
\cc{memcheck}.
\cc{initcheck} has the same abstraction issue on the temporal side.
Once any prior request has written a KV block, the underlying memory
is considered initialized.
An uninitialized read of the KV cache entry is therefore not reported by
\cc{initcheck}.

\subsection{Design Requirements}
\label{ss:motiv-requirements}

The case studies~(\autoref{ss:motiv-bugs}) and
limitations~(\autoref{ss:motiv-limitations}) together
motivate three design requirements for a practical KV cache safety
checker.
A checker must reason about serving-level abstractions that
traditional detectors cannot express (\autoref{ss:motiv-limitations}).
Then it must catch various violations~(\autoref{t:motiv-bug-list})
across the three boundaries that the case studies span:
CPU allocation, CPU-to-GPU translation, and GPU access.

\PP{R1. Early Detection}%
A KV cache violation should be caught close to the operation that
creates it,
not after corrupted outputs appear.
Case 2 and Case 3 both motivate this requirement.
In Case 2,
the wrong free of \cc{PB0} surfaces only when \cc{Req2},
an unrelated later request,
reads the overwritten KV vectors.
In Case 3,
\cc{Req1} reads uninitialized KV vectors from \cc{PB2} without any error,
and the prefix cache may later spread them to other requests.
In both cases, the corruption appears far from the operation that causes it.
The checker should therefore validate each step:
an allocation call when it is made,
the CPU block table before it reaches the GPU,
the GPU block table before a kernel consumes it,
and a kernel access at the access point itself.
Catching violations at the boundary that created them is
significantly helpful for triaging and fixing,
since it gives developers a precise failure site instead of a
downstream symptom whose root cause may live many requests away.

\PP{R2. Low Memory and Runtime Overhead}%
KV cache checks must preserve GPU memory capacity,
since KV cache size directly bounds batch size and context length.
This rules out directly porting mechanisms that buy detection by
reserving large extra memory,
such as ASan-style redzones, shadow memory, and quarantines,
as well as the allocation padding option provided by Compute
Sanitizer~\cite{nvidia-compute-sanitizer}
(\autoref{ss:motiv-limitations}).

This is not only a performance issue.
When a checker consumes excessive device memory, the serving system has to
shrink its KV cache, serve fewer concurrent requests, or may fail with CUDA
out-of-memory before the workload runs.
Compute Sanitizer documents that its tools can have a large memory footprint due
to tracking data, and that \cc{memcheck} padding increases device memory
pressure and can cause additional CUDA out-of-memory
errors~\cite{nvidia-compute-sanitizer}.
Thus, a KV cache checker should be designed to add only the necessary amount of
metadata and checks, and to avoid reserving large extra memory for redzones or
quarantines.

\PP{R3. Seamless Integration}%
KV cache checks should fit the existing PagedAttention workflow
without disturbing the single-buffer KV cache layout or the optimized
GPU kernels.
Cases 1 through 3 each manifest at a different boundary in this
workflow, so the checker should attach to those same boundaries:
the CPU allocator, the CPU-to-GPU block table commit, and the GPU
access path.

\section{Design}
\label{s:design}

\begin{figure*}[t]
\centering
\includegraphics[width=\textwidth]{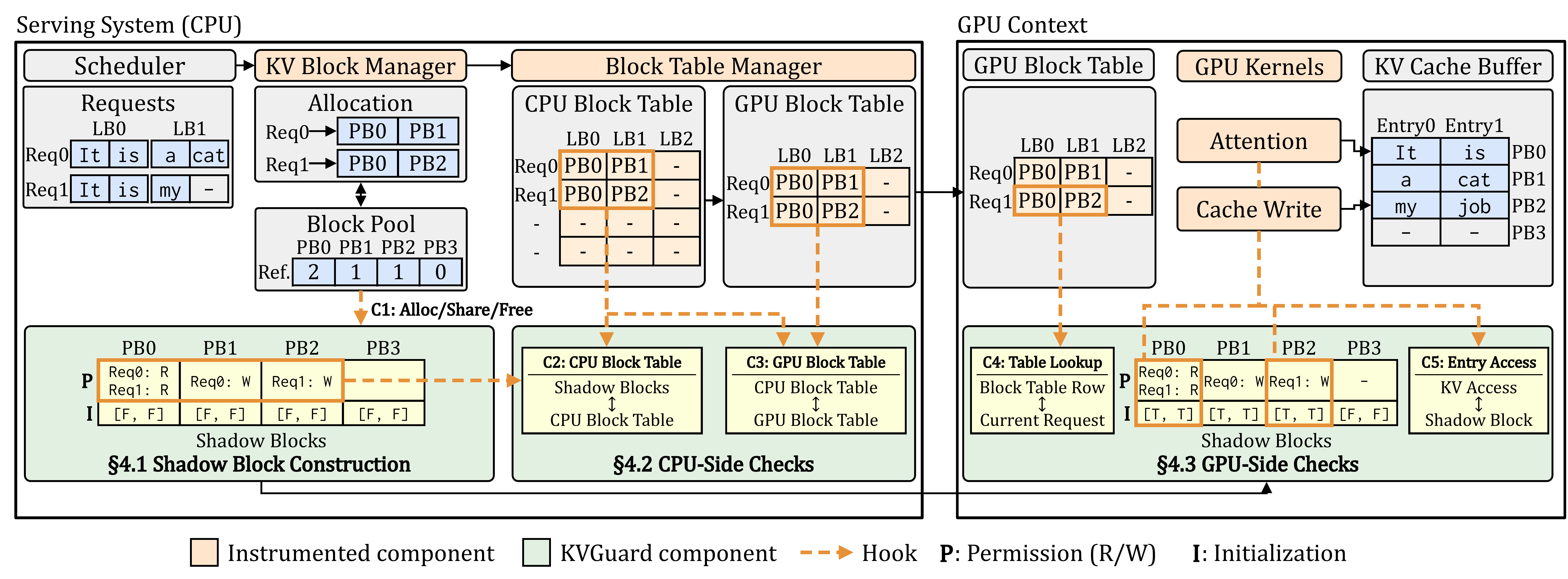}
\caption{Overview of \sys.}
\label{fig:design-overview}
\end{figure*}

\sys is a KV cache memory safety checker for LLM serving systems.
The key mechanism of \sys consists of two steps: maintaining shadow
metadata for critical objects in KV cache management, and checking the
serving system's behavior against the shadow metadata across the KV
cache management workflow.

\autoref{fig:design-overview} shows how \sys realizes this mechanism
across the PagedAttention workflow.
\sys first constructs and maintains shadow blocks
that record the safety state of each physical KV block,
by hooking CPU-side block allocation events
(\autoref{ss:design-shadow-metadata}).
\sys then validates the CPU-side state
that the serving system builds before each kernel launch,
including allocation calls, the CPU block table, and the GPU block table,
against the shadow blocks
(\autoref{ss:design-block-table-checks}).
Finally,
\sys validates each KV cache access inside the kernel
against the on-GPU shadow blocks
(\autoref{ss:design-gpu-runtime-checks}).

\subsection{Shadow Block Construction}
\label{ss:design-shadow-metadata}

\sys constructs and maintains a shadow block for each physical KV
block to track its safety state.
Each shadow block records two fields: permission and initialization.
These two fields together capture the ownership and content-validity
of a physical KV block along its lifetime, which are the properties
required to decide whether any given KV cache access is safe.
\sys updates these fields by hooking CPU-side allocation-related
events, and uses them to check the safety of operations related to KV
cache accesses.
This compact hook-based design satisfies R2 by avoiding redzones
around KV blocks, and R3 by attaching to existing allocation events
without changing the KV cache layout.

\PP{Permission}%
The permission field records
whether the request can read or write a given physical block during
the block's current lifetime.
For each block, this permission field is per-request metadata, meaning
that it tracks the permission of each request for the block.
In effect, the request ID acts as a lifetime token,
since the serving system never reuses an ID across requests.
Hence, this per-request encoding inherently captures the lifetime of
the block.
Specifically, when a block is freed and later reallocated to a new
request, the previous owner's permission is removed, and any new
permission is recorded under a different request ID.

Using this permission field, \sys detects three classes of unsafe
accesses.
In all three, the underlying check is the same:
the permission field of the target block does not record the required
access for the requesting request.
However, the three classes differ in why the record is missing.

% First, the target block is currently owned by a different request.
% %
% The permission field records the actual owner but not the requesting
% request, so the check fails.
% %
% Taking Case 2 in \autoref{ss:motiv-bugs} as an example,
% \cc{PB2}'s permission grants access to \cc{Req1},
% but \cc{Req0}'s kernel still attempts to read \cc{PB2}.
% %
% Second, the target block has never been allocated to the requesting
% request.
% %
% The permission field has no entry for the requesting request, so the
% check fails.
% %
% Taking Case 1 in \autoref{ss:motiv-bugs} as an example,
% the dummy run's \cc{Req0} writes to \cc{PB0},
% which was never granted to \cc{Req0}.
% %
% Third, the target block has been freed by its previous owner.
% %
% The previous owner's entry was removed on free, so the permission
% field no longer records any owner, and the check fails for the
% requesting request.
% %
% Taking Case 3 in \autoref{ss:motiv-bugs} as an example, \cc{PB1}'s
% only owner \cc{Req0} had already finished and released \cc{PB1}, but
% \cc{Req1}'s kernel still attempts to read \cc{PB1}.

First, the requesting request writes to a block that it should only read.
Because the block is shared,
the permission field records only read permission for the requesting request,
so the check for the write fails.
Taking Case 1 in \autoref{ss:motiv-bugs} as an example,
\cc{Req1} holds only read permission on the shared \cc{PB0},
but its cache write kernel still writes the KV vectors of "\cc{job}" into \cc{PB0}.
Second, the target block is not allocated to the requesting request.
The permission field records only the actual owners of the block, if any,
so the check fails for the requesting request.
For example, in B2 (\autoref{t:motiv-bug-list}),
padding rows of a CUDA graph write into blocks that belong to live requests.
Third, the target block has been freed and reused
after the requesting request acquired it.
The entry of the requesting request was removed on free,
so the permission field no longer records it,
and the check fails.
B0 and B10 in \autoref{t:motiv-bug-list} fall into this class.

\begin{figure*}[t]
\centering
\begin{subfigure}[t]{0.13\textwidth}
\centering
\includegraphics[width=\linewidth]{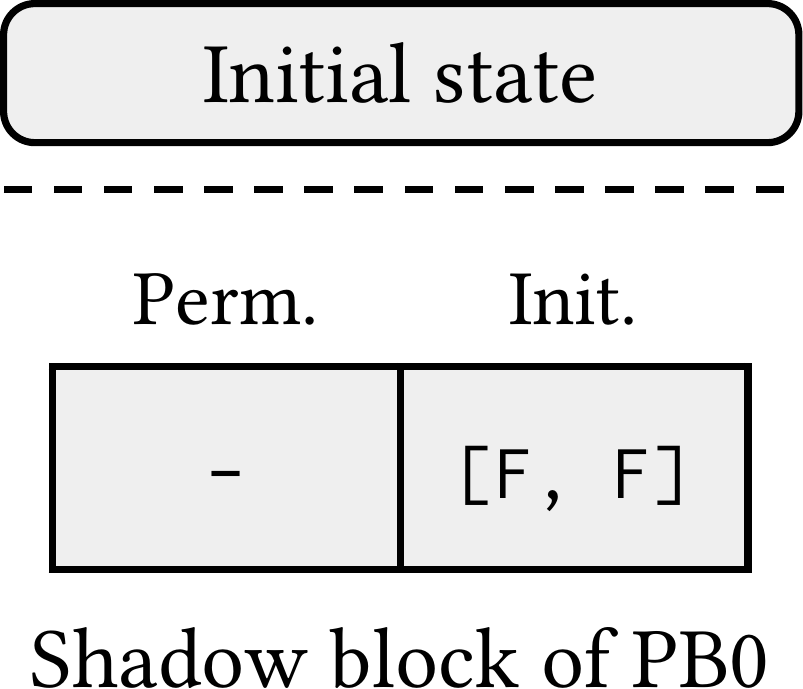}
\caption{Initial state.}
\end{subfigure}
\raisebox{1.2em}{%
\includegraphics[width=0.024\textwidth]{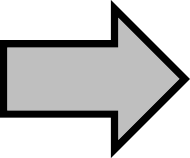}%
}
\begin{subfigure}[t]{0.13\textwidth}
\centering
\includegraphics[width=\linewidth]{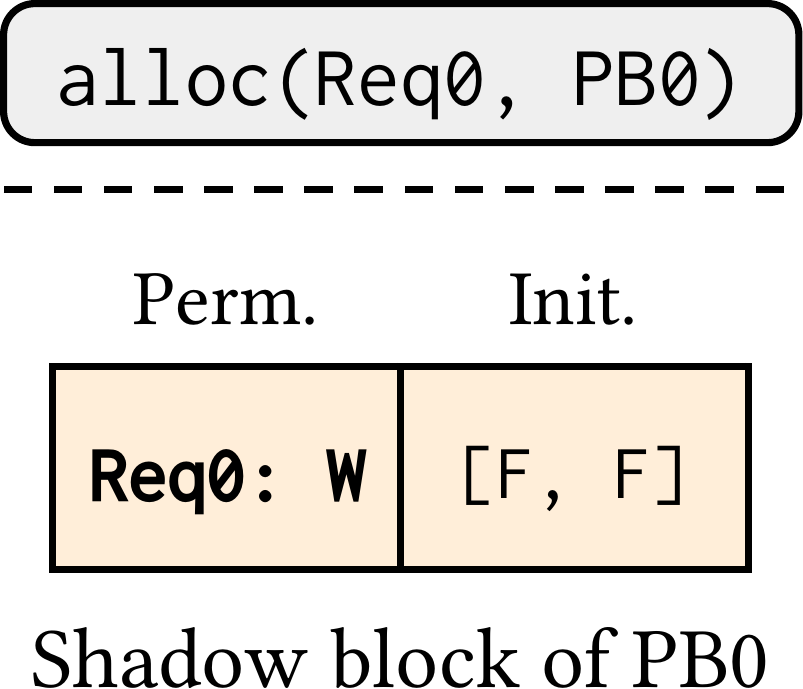}
\caption{Req0 allocation.}
\end{subfigure}
\raisebox{1.2em}{%
\includegraphics[width=0.024\textwidth]{fig/design/shadowblock_arrow.pdf}%
}
\begin{subfigure}[t]{0.13\textwidth}
\centering
\includegraphics[width=\linewidth]{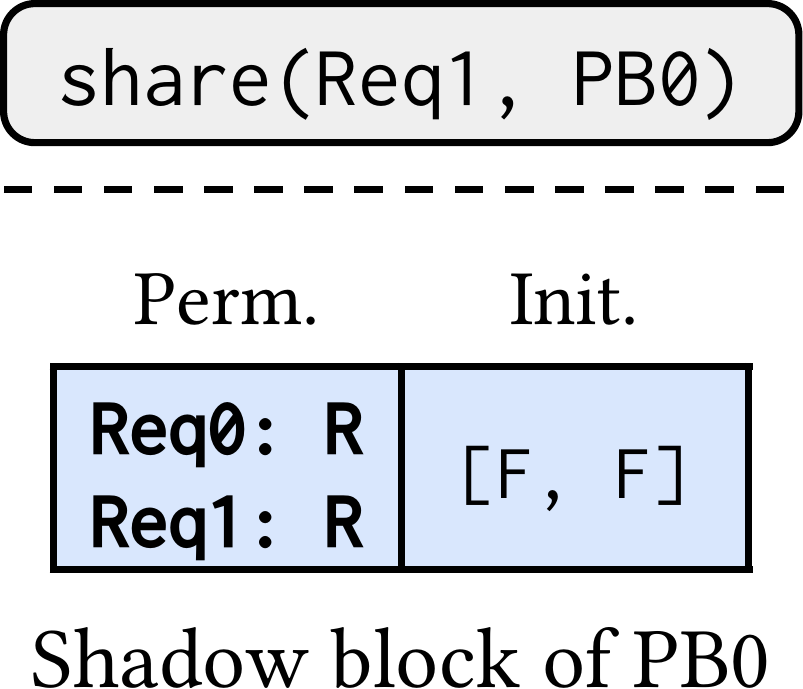}
\caption{Prefix sharing.}
\end{subfigure}
\raisebox{1.2em}{%
\includegraphics[width=0.024\textwidth]{fig/design/shadowblock_arrow.pdf}%
}
\begin{subfigure}[t]{0.13\textwidth}
\centering
\includegraphics[width=\linewidth]{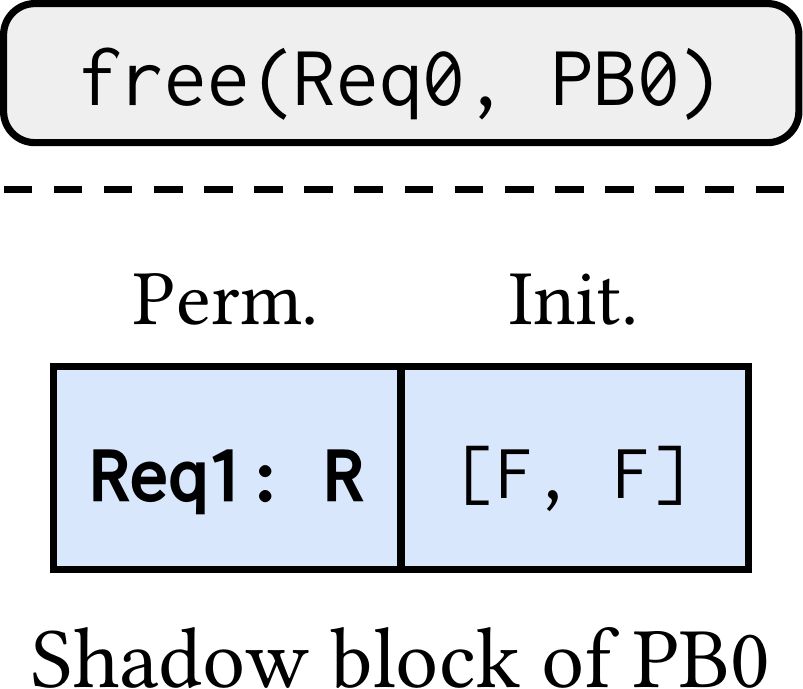}
\caption{Req0 ends.}
\end{subfigure}
\raisebox{1.2em}{%
\includegraphics[width=0.024\textwidth]{fig/design/shadowblock_arrow.pdf}%
}
\begin{subfigure}[t]{0.13\textwidth}
\centering
\includegraphics[width=\linewidth]{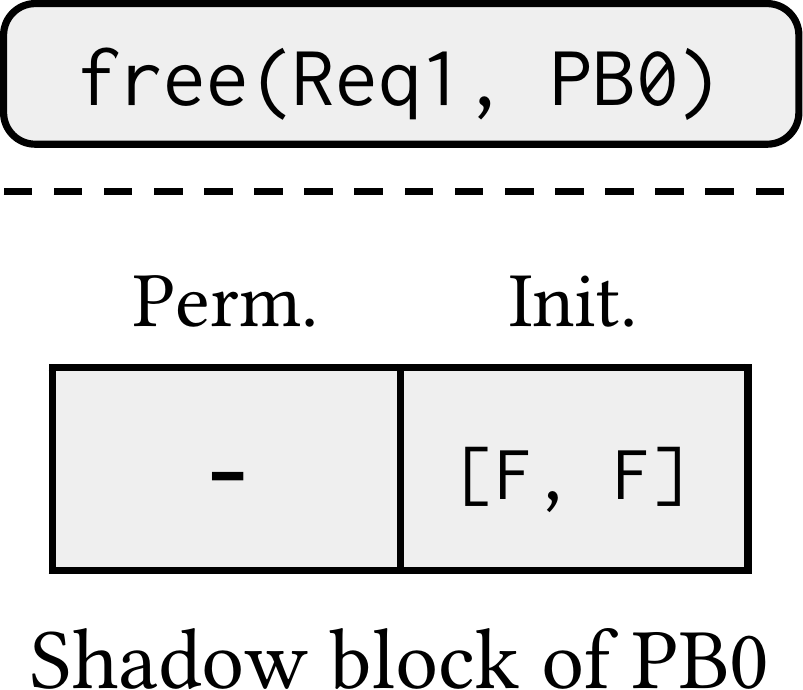}
\caption{Req1 ends.}
\end{subfigure}
\raisebox{1.2em}{%
\includegraphics[width=0.024\textwidth]{fig/design/shadowblock_arrow.pdf}%
}
\begin{subfigure}[t]{0.13\textwidth}
\centering
\includegraphics[width=\linewidth]{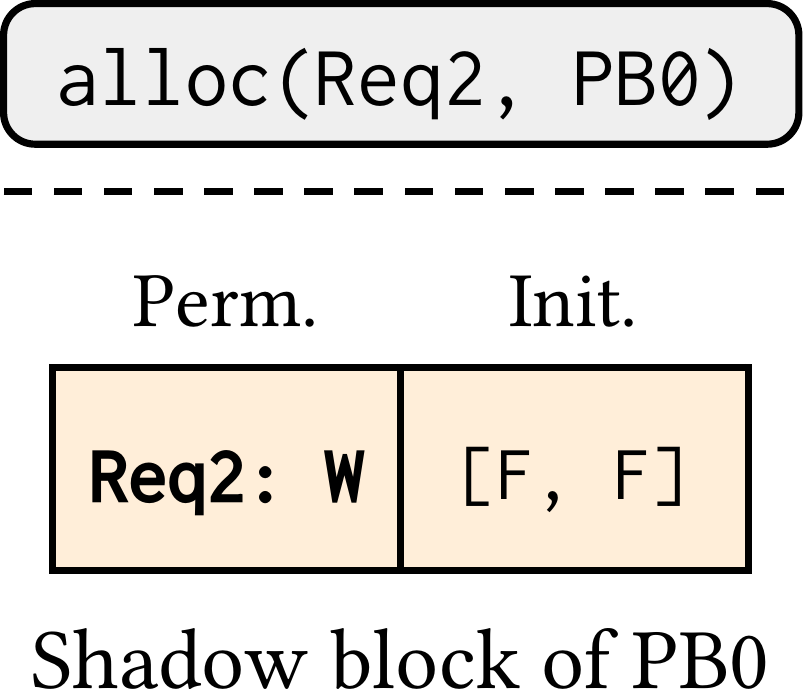}
\caption{Req2 allocation.}
\end{subfigure}
\caption{Shadow block updates across CPU allocation events.
Each panel shows the shadow block state of PB0.}
\label{fig:shadow-block-update}
\end{figure*}

\autoref{fig:shadow-block-update} shows how \sys updates
the permission of \cc{PB0} across requests.
%
% \JU{Add this single sentence below.}
Initially, \cc{PB0} is free and thus its permission is \cc{-}, meaning
that no permission is granted to any request.
On \cc{Req0}'s allocation of \cc{PB0}, \sys grants write permission to
\cc{Req0}, making \cc{PB0} writable only by \cc{Req0}.
On prefix sharing, \sys downgrades \cc{Req0}'s permission to read and
also grants read permission to \cc{Req1},
making \cc{PB0} read-only for both requests.
As \cc{PB0} is freed per request, \sys accordingly removes their
permissions one by one.
On \cc{Req2}'s later allocation of the same \cc{PB0},
\sys grants write permission to \cc{Req2};
any later kernel that still refers to \cc{Req0} or \cc{Req1} finds no
permission on \cc{PB0} and is rejected as a stale access.

\PP{Initialization}%
The initialization field tells whether each KV entry inside the block
is properly initialized.
It is worth noting that, while the permission field is per-block
metadata, the initialization field is per-entry metadata.
This per-entry granularity for initialization is needed because a
block can be partially written, and thus the remaining unwritten entries
hold stale data, which can be \cc{NaN}, garbage, or values written by a
previous request (e.g., B13 and B24
in~\autoref{t:motiv-bug-list}).
To track the initialization state, \sys sets the initialization bit on
each write, and clears the bits in a block when the block is
(re)allocated to a request.
This allows \sys to detect a read of an uninitialized entry that has not been
written, which is described later (\autoref{ss:design-gpu-runtime-checks}).

\subsection{CPU-Side Checks}
\label{ss:design-block-table-checks}

\sys performs three CPU-side safety checks that validate the
construction of CPU-side state.
The serving system builds this state in three sequential steps:
(i) allocation-related calls update the block pool,
(ii) a CPU block table is assembled from the allocated blocks,
and (iii) a GPU block table is derived from the CPU block table.
Following the same order, \sys places an individual check at each
step, as detailed below.
This step-by-step placement realizes R1 on the CPU side, since a wrong
allocation, CPU block table, or GPU block table is rejected at the
step that produced it, before the kernel ever launches.

\PP{C1. Checking Alloc/Share/Free Calls}%
% \sys checks the correctness of the block allocation decisions by the serving
% system.
% %
% Since the decisions can be wrong due to a bug in the serving system,
% \sys checks the exsisting metadata in the serving system: the reference counts
% of each physical block.
% %
% Specifically, \sys validates three allocation events: allocation,
% sharing, and free.
% %
% For allocation, \sys checks if the block is currently free, i.e., no
% the reference count is zero.
% %
% For sharing, \sys checks that the block is currently allocated, i.e.,
% the reference count is non-zero.
% %
% % Otherwise, the sharing attempt is a bug because the block is not
% % allocated to any request.
% %
% For free, \sys checks that the block is currently referenced by a single
% request, i.e., the reference count is one.
\sys checks the validity of the CPU-side block allocation-related
calls using shadow blocks.
%
% Since these call invocation alone can be wrong due to a bug in the
% serving system, \sys checks each allocation decision against the
% current shadow block state.
% %
Specifically, \sys validates three allocation events: allocation,
sharing, and free.
For allocation, \sys checks if the block is currently free, i.e., no
request has permission in the corresponding shadow block.
Otherwise, the allocation call is a bug because the block is still
in use by another request (e.g., B11 in~\autoref{t:motiv-bug-list}).
For sharing, \sys checks that the block is currently allocated, i.e.,
at least one request has permission.
Otherwise, the sharing call is a bug because the block should be
allocated before sharing.
For free, \sys checks that the block is currently referenced by a single
request, i.e., only the freeing request has permission.
Otherwise, the free call is a bug, leading to either double-free (if
no request has permission, e.g., B16) or
use-after-free (if multiple requests have permission, e.g., B17).
It is worth noting that, while \sys uses shadow blocks to check the
allocation decisions, those checks can also be performed using the
block pool's reference counts, because the two are updated in the same
way on the same events.
This means that this check can be seamlessly integrated into the
serving system.

\PP{C2. Checking CPU Block Table}%
% \BL{this uses both shadow block and shadow block table?}
% \JU{Updated.}
% %
\sys checks the safety of the CPU block table using shadow blocks.
Before launching a kernel, the serving system builds a CPU block
table, mapping logical block indices to physical block indices for
each request.
To ensure the safety of this mapping construction, \sys iterates over
every entry in the request's CPU block table.
For each entry, \sys first obtains the mapped physical block index.
Then \sys obtains the shadow block's permission field corresponding to
the physical block index, and checks if the permission grants the
required access to the request.

% % \BL{i don't think this step is necessary? this steps looks straightforward,
% % making shadow block table looks pointless}
% \BL{check this}
% \JU{Updated.}
%

For instance, as shown in \autoref{fig:design-overview}, if Req0's CPU block
table maps LB0 to PB0, then \sys checks whether the shadow block of PB0 grants
read or write permission to Req0.
On the other hand, if Req0's CPU block table maps LB1
to PB1, which is a stale mapping, then \sys checks the shadow block of PB1 and
detects that it does not grant permission to Req0.

% The corresponding shadow block must grant the required permission to the
% request: read permission for attention reads and write permission for cache
% writes.
% %
% For example, if Req0's block table entry maps a physical block PB0
% that belongs only to Req1, \sys detects it by checking the permission
% of shadow block of PB0.

\PP{C3. Checking GPU Block Table}%
\sys checks if the GPU block table (to be transferred to GPU memory)
is correctly built from the CPU block table.
%
% For every request and logical block index of the request, \sys
% compares the GPU block table entry with the corresponding CPU block
% table entry.
%
To this end, for every request row in the GPU block table, \sys checks
the following constraints: \cc{GPUBlockTable[req][LB]} is the same as
\cc{CPUBlockTable[req][LB]}.
This comparison catches a wrong or stale mapping that reaches the GPU
block table under asynchronous or overlap scheduling (e.g., B1 and B8
in~\autoref{t:motiv-bug-list}), as well as failures in the copy from
CPU state to GPU state.

\subsection{GPU-Side Checks}
\label{ss:design-gpu-runtime-checks}

After CPU-side checks, \sys further checks the safety of the KV cache
access inside GPU kernels.
To support these GPU-side checks, \sys first copies the shadow blocks
to GPU memory before kernel launch.
Then it performs two different checks using the shadow blocks: GPU
block table lookups (C4) and KV cache entry accesses (C5).
These in-kernel checks complete R1 by reporting a violation at the
access point itself, rather than as a downstream output corruption
many requests later.

% \BL{this sentence is strange. it's detail. this sentence introduce C4
% and C5. this only covers C4?} \BL{do not mention specific bugs here.
% mention those inside C4 and C5 descriptions} \JU{Updated.}
% Even if the CPU and GPU block tables were valid from CPU-side checks,
% the kernel can still violate the KV cache safety: (i) the kernel looks
% up the GPU block table entry that the request should not use (\BL{which
% bugs} \JU{Updated} B3 in \autoref{t:motiv-bug-list}); or (ii) the
% kernel computes a wrong entry location inside the block (B4 and B5 in
% \autoref{t:motiv-bug-list}).
%
% To detect these cases, \sys copies the shadow blocks to GPU memory
% before kernel launch, and instruments GPU-side KV cache accesses to
% validate them against the on-GPU shadow metadata.

\PP{C4. Checking the GPU Block Table Lookup}%
When the kernel looks up a GPU block table to locate the physical
block for a logical block, \sys checks if the GPU block table entry is
a valid one for the current request.
A valid lookup must satisfy two conditions: the entry must belong to the current
request's block table row, and the logical block index must be within the
entries that the request is allowed to use.
%
% \BL{say we use shadow blocks to do this?}
% \JU{Updated.}
To achieve this, \sys reuses the block table bounds checked before
launch.
At runtime, when a kernel reads a GPU block table entry, \sys checks that the
lookup is within the valid entries of the current request's row.
This catches real-world bugs such as B19
(\autoref{t:motiv-bug-list}).

% For this, \sys records which GPU block table entries are allowed for each
% request before launch, and checks each GPU block table lookup against this
% allowlist at runtime.
%

\PP{C5. Checking the KV Cache Entry Access}%
Lastly, to ensure the safety of the actual KV cache access, \sys
checks if the current request is allowed to access the target physical
block (and its entry) using the shadow blocks.
Specifically,
\sys first locates the shadow block corresponding to the target
physical block index,
and then performs two checks against it:
permission and initialization.
For permission,
\sys checks that the permission field of the shadow block grants the
required access to the current request,
where a read access requires the read permission and a write access
requires the write permission.
%
% This check catches two class of bugs: (i) accesses to a block that
% belongs to another request (e.g., B2, B3, and B7
% in~\autoref{t:motiv-bug-list}), and (ii) stale accesses to a block
% that has been freed and reused after the request acquired it (e.g.,
% B0, B10, and N5), because the previous owner's permission is removed
% once the block is reallocated to another request.
% %
This check catches three classes of bugs:
(i) writes to a shared, read-only block (e.g., B3 in~\autoref{t:motiv-bug-list}),
(ii) accesses to a block that belongs to another request (e.g., B2 and B7),
and (iii) stale accesses to a block that has been freed and reused
after the request acquired it (e.g., B0, B10, and N5),
because the previous owner's permission is removed
once the block is reallocated to another request.
For initialization, \sys checks that the initialization field of the
target entry is set before a read, and sets the initialization field
on a write.
This check ensures that a read never observes an uninitialized entry
that has not been written within the current block lifetime (e.g., B6,
B15, and N3).

\begin{figure}[t]
\centering
\tiny
\input{code/kvguard_access.cu.tex}
\caption{\sys instrumentation of GPU KV cache access.}
\label{fig:design-access-pseudocode}
\end{figure}

\autoref{fig:design-access-pseudocode} shows the \sys-instrumented CUDA cache
access path.
\sys checks the GPU block table entry,
validates the physical block's permission through the shadow block,
and checks or stamps the entry initialization around the KV cache access.

\subsection{KV Transfer Checks}%
\label{ss:design-kv-transfer}
KV offloading~\cite{gao2024cost,qin2026mooncake}
and P/D disaggregation~\cite{zhong2024distserve}
move KV blocks through DMA and the network.
\sys handles such cases by treating each transfer as an access by the request on
whose behalf it runs.
On the sender, a transfer is a read:
its source entries must be initialized (C5),
and its source blocks stay held by the request until the transfer completes,
so a free before completion is a use-after-free (C1).
On the receiver, a transfer is a write:
its destination blocks must be those of the receiving request (C3),
they stay held until completion (C1),
and an entry becomes initialized only when the transfer actually writes it (C5).

A transfer may also deliver fewer entries than the receiver reserved.
\sys marks only the entries delivered as initialized, so a later read access to
an entry that was not delivered is caught as an uninitialized read (C5).
Here, the receiver cannot tell a short delivery from a full one by itself,
because both complete the same way.
The sender therefore reports the count of entries it wrote.
This count is the only metadata \sys adds to a transfer,
and each node still checks against its own shadow blocks.
B17, B21, N4, and N7 in \autoref{t:motiv-bug-list} are caught this way.

\section{Implementation}
\label{s:impl}

\sys is implemented on vLLM~\cite{vllm} and SGLang~\cite{sglang},
two popular open-source LLM serving systems
that use \PgdAttn for KV cache management.
The implementation has three parts for each system:
(i) a Python package
that manages the shadow blocks, CPU-side bridge, and Triton check helpers,
(ii) CUDA check kernels and helpers
that are included from the system's cache and attention kernels,
and (iii) hook insertions
across the system's block manager, block table, attention backends, and
kernels.
\autoref{t:impl-loc} lists their sizes:
lines exclude blanks, comments, and tests.
% and the hook insertions count the paths the evaluation exercises,
% leaving out the hooks for sparse attention, hybrid state, disaggregation,
% host offload, and speculative decoding.

\begin{table}[t]
\centering
\caption{Implementation size in lines of code.}
\label{t:impl-loc}
\scriptsize
\begin{tabular*}{\columnwidth}{@{\extracolsep{\fill}}lrr@{}}
\toprule
Component & vLLM & SGLang \\
\midrule
Shadow blocks and checks (Python) & 5,730 & 18,419 \\
GPU check kernels and helpers (CUDA, C++) & 385 & 3,285 \\
Hook insertions in the system and its kernels & 1,663 (180 sites) & 1,535 (177 sites) \\
\bottomrule
\end{tabular*}
\end{table}

\PP{Shadow Block Construction}%
\sys maintains one shadow record per physical block,
carrying the two fields described in \autoref{ss:design-shadow-metadata}.
The permission field is stored as a per-request permission list,
recording which requests currently hold the block
and whether each holds it for reading or writing.
The initialization field is stored as a per-entry bit,
set when a kernel writes the entry
and cleared in bulk when the block is (re)allocated to a request.

To keep the shadow records consistent with vLLM's block lifecycle,
\sys intercepts five state transitions in vLLM's block pool and KV cache manager:
allocation, touch, cache promotion, free, and cached eviction.
These cover both block-commit paths in vLLM v1,
explicit allocation and prefix-cache promotion,
without duplicating bookkeeping.

\PP{CPU-Side Checks}%
CPU-side checks are fused with the hook points
that manage the relevant state.
C1 runs inside the allocation hooks themselves,
so an allocator event raises immediately on a wrong precondition
and adds at most one integer test per event.
C2 and C3 share a single entry point invoked once before every kernel launch,
operating on a per-request block table snapshot
captured when the block table manager commits the block table.
C2 first checks that the kernel's effective row length
does not exceed the committed row length in the snapshot.
For each cell within that bound,
C2 then verifies that the cell's physical block is held by the request,
and rejects the launch on a mismatch.
C3 then compares the snapshot against the GPU block-table tensor cell by cell.

\PP{GPU-Side Checks}%
GPU-side checks are realized as kernel instrumentation.
\sys exposes a KV read check helper and a KV write check helper,
inlined into every GPU kernel that reads or writes the KV cache,
including attention, cache copy, cache write, gather, and SSM kernels.
The instrumented kernel accesses the shadow metadata
on KV access paths during runtime.
Before each scheduling step,
\sys packs the host shadow blocks into fixed-shape tensors,
including a row-to-request map, a host-visible violation counter, and a
violation ring buffer.
Each instrumented kernel reads the row-to-request map
to find the owning request of its batch row.
It then invokes the check helper with the request ID
and the looked-up physical block
to realize C4 and C5.
A cache write sets the initialization bit of the accessed entry, and a cache
read accepts the entry only when its bit is set.
\sys reports device-side violations in two modes.
The aborting mode stops on the first violation and prints a decoded report,
keeping it close to the root cause.
The logging mode records violations in a ring buffer drained periodically by
the host, letting one run collect multiple reports for debugging.

\PP{KV Transfers Without a Kernel}%
As described in \autoref{ss:design-kv-transfer}, KV transfers are checked on the
host: \sys hooks the submission of each offload copy and P/D transfer in the
systems' connectors and its completion event, and checks the source or
destination blocks against the shadow blocks.
For a cross-node P/D transfer, the sender records the number of entries it wrote
and passes that count to the receiver, which initializes only that many
destination entries, so a short delivery leaves the rest uninitialized.

% \PP{Operation Modes}%
% \sys exposes three operation modes
% that trade coverage for runtime cost:
% \cc{C1-only}, \cc{C1-C3}, and \cc{Full}.
% %
% \cc{C1-only} keeps only the allocator-side checks
% fused into the block manager,
% catching alloc/share/free precondition violations
% with no per-launch or per-access overhead.
% %
% \cc{C1-C3} additionally runs the pre-launch block-table checks,
% leaving GPU kernels uninstrumented.
% %
% \cc{Full} additionally enables the GPU-side checks (C4 and C5).

\section{Evaluation}
\label{s:eval}

This section evaluates whether \sys can check KV cache safety at acceptable
serving cost.
The evaluation answers three questions:
(i) how much runtime overhead \sys adds to serving systems,
(ii) how much memory \sys uses for shadow metadata,
% (iii) whether the fail-fast checker runs cleanly on safe workloads,
and (iii) whether the checks detect real and injected KV cache safety bugs.

\subsection{Experimental Setup}
\label{ss:eval-setup}

\PP{Platform}%
All experiments run on a server with one NVIDIA H100 80\GB GPU and an Intel
Xeon Platinum 8468 processor (16 vCPUs).
The software stack uses Ubuntu 24.04, CUDA 13.0, vLLM v0.32.0 with
FlashAttention 3~\cite{shah2024flashattention}, and SGLang v0.5.22 with
FlashInfer 0.7.1.
\sys attaches through each engine's plugin mechanism without patching either.

\PP{Workloads}%
\autoref{t:eval-workloads} lists the evaluated models.
Llama 3.2 1B~\cite{grattafiori2024llama} stresses sanitizer overhead
because each decode step has little model compute,
while Qwen 2.5 14B~\cite{yang2024qwen25} represents a larger model
where KV and attention work dominate each step.
The serving workload uses ShareGPT V3 prompts~\cite{sharegpt52k} on every model.
Llama 3.2 1B runs two more datasets:
(1) LongBench~\cite{bai2024longbench} prompts of 2K to 8K tokens,
to evaluate long inference sequences,
and (2) five-shot MMLU~\cite{hendrycks2021measuring} over eight subjects
to exercise prefix publication and sharing.
%
% vLLM runs at its default GPU memory utilization of 0.90 and SGLang at a
% static memory fraction of 0.85; both engines receive the same KV pool per
% model (\autoref{ss:eval-memory}), so these fractions only bound the pool,
% except in the Compute Sanitizer runs (\autoref{ss:eval-performance}).
% %

\PP{Configurations}%
The evaluation uses two configurations.
\cc{Baseline} runs the serving systems with \sys disabled,
and \cc{KVGuard} runs them with \sys attached and every check (C1 to C5)
enabled.
\cc{KVGuard} is evaluated in the aborting mode described in
\autoref{s:impl}.

\begin{table}[t]
\centering
\caption{Evaluation workloads.}
\label{t:eval-workloads}
\scriptsize
\begin{tabular*}{\columnwidth}{@{\extracolsep{\fill}}lrll@{}}
\toprule
Model &
Size &
Role &
Dataset \\
\midrule
Llama 3.2 &
1B &
Smallest (worst case) &
ShareGPT V3, LongBench, MMLU \\
Llama 3.1 &
8B &
Mid-size &
ShareGPT V3 \\
Qwen 2.5 &
14B &
Largest (compute-bound) &
ShareGPT V3 \\
Qwen 3.5 &
4B &
Hybrid (recurrent state) &
ShareGPT V3 \\
\bottomrule
\end{tabular*}
\end{table}

% On the first GPU-side violation,
% \sys flushes the violation record to host-visible memory,
% traps the kernel,
% and exits the serving process after printing the decoded report.
%
% This setting matches the default \cc{Full} mode used by the implementation.
%
% Thus, we perform evaluation with a minimal patch for N2: padded entries touched
% by the evaluated FA3 path are marked consistently with the intended mask
% behavior.
%
% Without this patch,
% \cc{Full} would correctly stop on the known N2 finding
% instead of measuring normal serving overhead.

\subsection{Performance Overhead}
\label{ss:eval-performance}

\begin{figure*}[t]
\centering
\includegraphics[width=\textwidth]{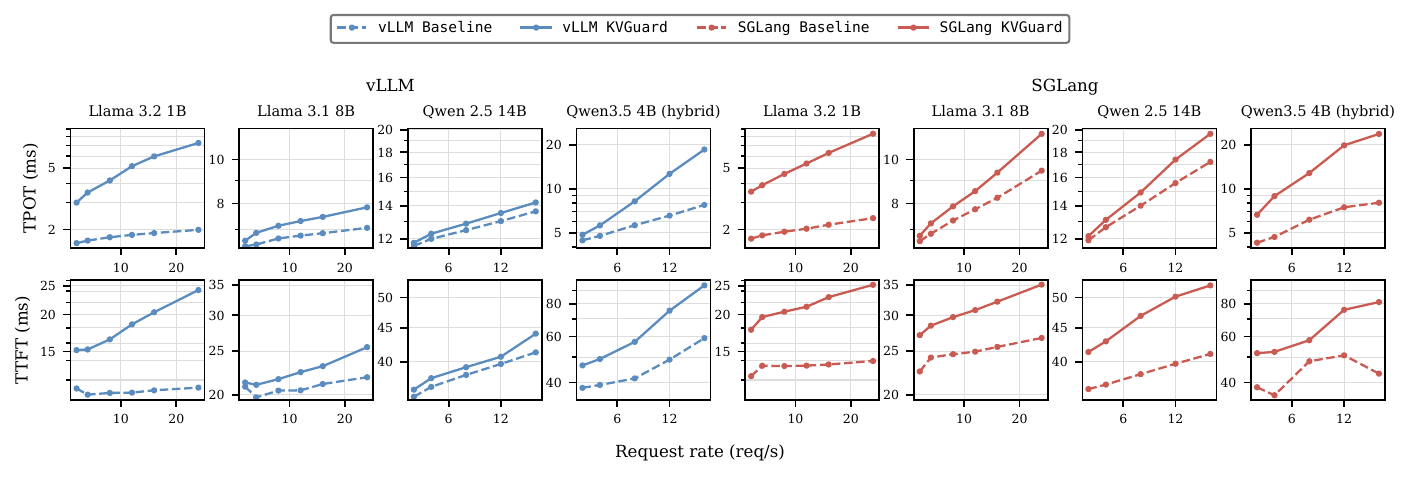}
\caption{Median TPOT and TTFT of \cc{Baseline} and \sys across request rates.}
\Description{Two rows of eight panels: median TPOT and TTFT per model, one block of panels per engine.}
\label{fig:eval-latency}
\label{fig:eval-tpot-breakdown}
\end{figure*}

\sys's overhead is largest on the smallest model and shrinks as the model's
size grows.
\autoref{fig:eval-latency} shows the median TPOT and TTFT across request
rates on the four models and both serving systems.
On Llama 3.2 1B, the TPOT slowdown grows from 1.83\x at 2 req/s to 3.67\x at
24 req/s on vLLM and from 2.02\x to 3.53\x on SGLang.
This is the worst case in the evaluated matrix:
the model's baseline TPOT is only 1.6 to 1.7 ms,
where the per-step host work for the shadow is as large as the execution time
of the step itself.
On Llama 3.1 8B, the slowdown is 1.03\x to 1.11\x on vLLM and 1.03\x to
1.21\x on SGLang,
and on Qwen 2.5 14B 1.01\x to 1.04\x and 1.02\x to 1.14\x,
because each step already spends 6 to 17 ms in model execution.
The hybrid Qwen 3.5 4B sits between, at 1.09\x to 2.39\x and 1.55\x to
2.95\x,
because its recurrent-state pool adds a state dispatch to every step besides
the attention ones.
TTFT follows the same order, from 1.35\x to 2.14\x on Llama 3.2 1B down to
1.02\x to 1.27\x on Qwen 2.5 14B,
and stays below the TPOT slowdown because much of the time to first token is
spent outside the checked step.
SGLang shows the larger slowdown on the two larger models because its overlap scheduler
hides its own per-step host work behind the GPU,
so the attachment's host work is exposed in full.

\begin{figure}[t]
\centering
\includegraphics[width=\columnwidth]{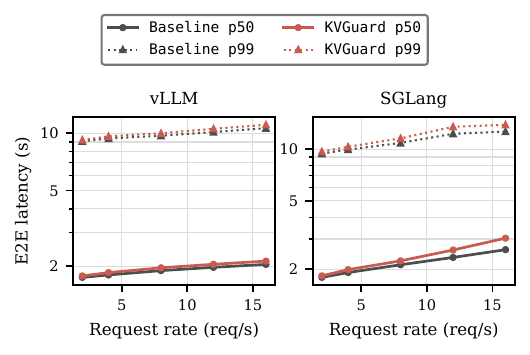}
\caption{End-to-end latency of \cc{Baseline} and \sys on Qwen 2.5 14B.}
\Description{Two-panel line chart of median and p99 end-to-end latency on vLLM and SGLang.}
\label{fig:eval-e2e}
\label{fig:eval-qwen-latency}
\end{figure}

\begin{figure}[t]
\centering
\includegraphics[width=\columnwidth]{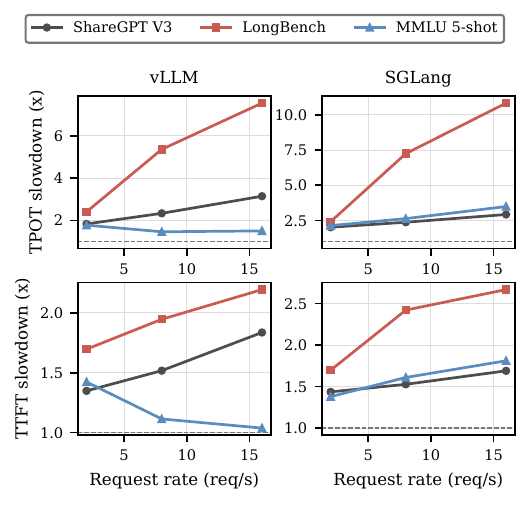}
\caption{TPOT and TTFT slowdown of \sys per workload on Llama 3.2 1B.}
\Description{Four-panel line chart of slowdown per workload on vLLM and SGLang.}
\label{fig:eval-workloads}
\end{figure}

\sys has moderate user-visible latency impact
on the larger evaluated model.
\autoref{fig:eval-qwen-latency} shows the end-to-end latency
for Qwen 2.5 14B.
At 16 req/s,
\sys increases median end-to-end latency from 2.04 s to 2.13 s on vLLM
and from 2.60 s to 3.04 s on SGLang,
and the p99 from 10.6 s to 11.1 s and from 12.7 s to 13.8 s.
This result shows that \sys remains feasible
on the larger model even under high request pressure.
The tail cost falls on the small model instead:
on Llama 3.2 1B at 24 req/s,
the p99 TPOT rises from 2.4 ms to 17.3 ms on vLLM
and from 4.0 ms to 17.4 ms on SGLang.

The overhead also depends on the prompt structure of the workload.
\autoref{fig:eval-workloads} shows the TPOT and TTFT slowdown of \sys per
workload on Llama 3.2 1B.
LongBench prompts of 2K to 8K tokens raise the TPOT slowdown to 7.55\x on
vLLM and 10.84\x on SGLang at 16 req/s,
because the dispatch check covers every position a step consumes
and a long context multiplies them.
Five-shot MMLU, whose prompts share a prefix of 0.7K to 2.4K tokens per
subject, costs 1.46\x to 1.78\x on vLLM and 2.13\x to 3.49\x on SGLang,
close to ShareGPT,
because the shared prefix is published once and then only referenced.
On vLLM the MMLU slowdown even falls with the rate,
because the detached engine's own decode step slows down on the 2K-token
contexts while \sys's cost per step grows little.

\begin{figure*}[t]
\centering
\includegraphics[width=\textwidth]{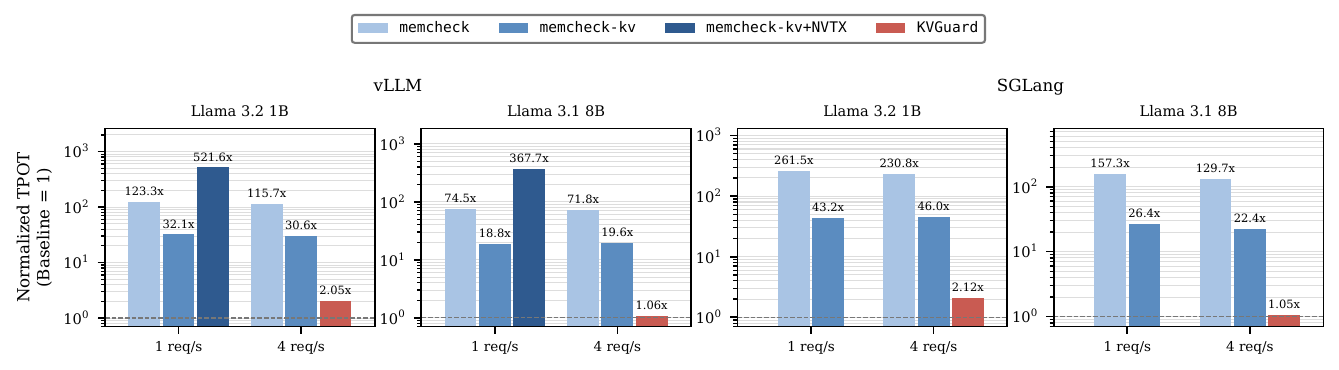}
\caption{TPOT slowdown of \sys and NVIDIA Compute Sanitizer, normalized to \cc{Baseline}.}
\Description{Four-panel log-scale bar chart of KVGuard and memcheck slowdown.}
\label{fig:eval-compute-sanitizer}
\end{figure*}

\PP{Comparison with Compute Sanitizer}%
\sys's overhead is much lower than that of NVIDIA Compute
Sanitizer~\cite{nvidia-compute-sanitizer}
(\autoref{fig:eval-compute-sanitizer}).
The comparison uses Compute Sanitizer \cc{memcheck}, which is the closest
available GPU-side baseline for out-of-bounds, misaligned, and use-after-free
checks on CUDA allocations, in two forms.
\cc{memcheck} as shipped instruments every kernel of the serving process.
\cc{memcheck-kv} restricts the instrumentation to the KV cache kernels,
i.e., the attention and cache-write kernels that read or write the KV pool
(FlashAttention and \cc{reshape\_and\_cache} on vLLM,
FlashInfer's paged kernels and the pool's \cc{index\_put} on SGLang),
through the tool's kernel-name filter,
with host stack traces off,
so that the sanitizer checks the same kernels as \sys at the least cost the
tool allows.
At 4 req/s on Llama 3.2 1B, \sys raises the median TPOT of vLLM from 1.7 ms
to 3.5 ms (2.1\x),
whereas \cc{memcheck} raises it to 194 ms (115.7\x)
and \cc{memcheck-kv} to 51 ms (30.6\x).
On SGLang the same points are 2.1\x, 230.8\x, and 46.0\x.
On Llama 3.1 8B, \sys costs 1.1\x on both systems,
where \cc{memcheck} costs 71.8\x on vLLM and 129.7\x on SGLang
and \cc{memcheck-kv} 19.6\x and 22.4\x.

Here, both \cc{memcheck} and \cc{memcheck-kv} still check the KV pool as one
allocation, so an access to the wrong block inside the pool is invisible to
them.
Registering every KV block as a sub-allocation with the NVTX memory
API~\cite{nvtx} gives \cc{memcheck-kv} block granularity,
at 522\x on Llama 3.2 1B and 368\x on Llama 3.1 8B at 1 req/s on vLLM,
and at 4 req/s the engine under the tool slows to about one step per second,
so that point is not measured.
Thus \cc{memcheck-kv} remains 15\x to 22\x slower than \sys on the very
kernels \sys checks,
because it checks every memory access of those kernels,
while \sys checks each step's tables once before the kernels run.

Compute Sanitizer also required a more constrained setup than \sys.
The \cc{memcheck} runs set the GPU memory utilization of both engines to 0.50
and the KV pool to a quarter of its size,
because the sanitizer's internal shadow allocator runs out of memory at the
ordinary settings,
and they measure 40 prompts at 1 and 4 req/s.

In addition, Compute Sanitizer \cc{initcheck} is excluded because the systems
failed to run with \cc{initcheck} enabled.

\subsection{Memory Overhead}
\label{ss:eval-memory}

\autoref{t:eval-memory} shows the memory overhead of \sys shadow metadata.
\sys keeps shadow metadata on both GPU and host,
and the table reports its size relative to the two main memory consumers in
serving: the KV cache and model weights.
The main GPU component is the per-entry initialization record, whose size scales
with the number of physical blocks, the block size, and the number of layers.
Other GPU metadata, including per-block permission state, request-row mappings,
the violation counter, and the violation ring buffer, is comparatively small.
Host shadow memory is smaller still, because the host keeps one compact record
per physical block.

\begin{table}[t]
\centering
\caption{Memory overhead of \sys shadow metadata on H100.}
\label{t:eval-memory}
\scriptsize
\setlength{\tabcolsep}{2pt}
\begin{tabular}{@{}lccccc@{}}
\toprule
Model &
\makecell{KV\\Cache} &
\makecell{GPU\\Shadow} &
\makecell{Host\\Shadow} &
\makecell{Shadow /\\KV} &
\makecell{Shadow /\\Weights} \\
\midrule
Llama 3.2 1B &
32 GiB &
17.8 MiB &
17.2 MiB &
0.11\% &
1.5\% \\
Qwen 3.5 4B &
59 GiB &
23.2 MiB &
18 MiB &
0.07\% &
0.46\% \\
Llama 3.1 8B &
24 GiB &
9.6 MiB &
6.6 MiB &
0.07\% &
0.11\% \\
Qwen 2.5 14B &
24 GiB &
9.1 MiB &
4.4 MiB &
0.05\% &
0.05\% \\
\bottomrule
\end{tabular}
\end{table}

\autoref{t:eval-memory} shows that the shadow is 0.11\% of the KV cache
for Llama 3.2 1B
and 0.05\% for Qwen 2.5 14B,
and at most 1.5\% of the model weights.
The absolute KV cache sizes are set by the evaluation:
both engines are given the same KV pool per model,
1,048,576 tokens on Llama 3.2 1B, 196,608 on Llama 3.1 8B,
and 131,072 on Qwen 2.5 14B,
within the memory left beside the weights,
while the hybrid model's pool is sized by each engine.
This makes Llama 3.2 1B's KV pool the largest in slots on the same 80\GB GPU,
even though the model is the smallest.
The shadow overhead therefore follows the actual KV pool geometry,
not only the model parameter count.
For example, Llama 3.2 1B carries the largest shadow of the four models because
it has the most slots per gigabyte of KV cache.
%

% \subsection{Correctness Validation}
% \label{ss:eval-correctness}

% Finally, \sys does not report violations on the evaluated safe workloads.
% %
% The final performance sweep contains 72 clean-run cells:
% two models,
% two configurations,
% six request rates,
% and three repeats.
% %
% All \cc{Full} cells complete without a fail-fast abort.
% %
% A separate 50-prompt smoke test for Llama 3.2 1B
% also completes with zero aborts after the N2 patch.
% %
% These results show that the final implementation can run the checked paths
% used in the performance evaluation
% without spurious violation reports.

\subsection{Bug Detection Capability}
\label{ss:eval-bug-detection}

\sys detects every bug in \autoref{t:motiv-bug-list}:
the 29 historical bugs from the vLLM and SGLang patch histories
and the 8 newly discovered bugs, each reproduced end to end.
The historical bugs test whether \sys catches known failures
whose fixes were written by other developers.
The new findings test whether the same checks expose previously unknown
unsafe KV cache behavior,
and \autoref{t:eval-new-findings} lists them with the fix pull requests
submitted upstream.

\begin{table}[t]
\centering
\caption{New KV cache safety issues found by \sys.}
\label{t:eval-new-findings}
\scriptsize
\begin{tabular}{@{}llll@{}}
\toprule
ID &
Fix PR &
\sys Check &
Finding \\
\midrule
N0 &
\#42276 &
C2 &
Stale mapping after SWA eviction. \\
N1 &
\#62871 &
C1 &
Block reuse after a rejected stream chunk. \\
N2 &
\#64649 &
C1 &
Async KV load stalls with free blocks. \\
N3 &
\#45604 &
C5 &
Uninitialized state read at int8 checkpoint. \\
N4 &
\#45605 &
C3 &
Parallel samples share one P/D transfer. \\
N5 &
\#47390 &
C5 &
Finished request's freed row relaunched. \\
N6 &
\#47392 &
C2 &
Session state past its kept tokens. \\
N7 &
\#47250 &
C1 &
Aborted prefill freed during KV transfer. \\
\bottomrule
\end{tabular}
\end{table}

\PP{End-to-End Reproduction}%
Each historical bug is replayed on the exact buggy revision,
i.e., the parent of its fix commit,
and on the exact fix,
with a serving workload that triggers the bug.
\sys reports the violation on every buggy run and reports nothing on the
fixed run.
For example, B2 is a stale slot-mapping tail that chunked prefill leaves
behind for the padding rows of a full CUDA graph.
On the buggy revision, \sys reports 276 accesses to another request's block
through the padding rows in each run (C3),
and 30 of 64 outputs differ from the fixed revision.
B8 is a sequence-length tensor that the overlap worker keeps sharing with
later scheduling,
so attention consumes block mappings under bounds changed for a different
batch.
\sys reports the reuse of the in-flight tensor (C3),
and the two buggy runs produce outputs that differ from the fixed runs on 8 and
6 of 8 requests and from each other on all 8,
a silent corruption without any error.
Most historical bugs show no symptom at all in the replay.
B7 and B9, for instance, complete every request with the same output as the
fix,
and only the \sys report shows that a padding row wrote a live
sliding-window entry (B7)
or that a batch launched before its block mappings were produced (B9).
\sys therefore reports each bug at the unsafe transition itself,
before a silent output corruption happens.

\PP{New Findings}%
\autoref{t:eval-new-findings} lists the eight new bugs \sys reported during
end-to-end serving,
three in vLLM and five in SGLang.
Three are lifecycle bugs caught by C1.
N1 admits the end-of-stream sentinel of a rejected streaming input as a new
request while the first instance's blocks are still held by its outstanding
dispatch.
N2 rejects an asynchronous KV load whose admission margin for a speculative
step exceeds what the request can still use near the context limit,
so the request waits forever with free blocks available.
N7 releases the KV of a cancelled prefill while the decode side's transfer
still reads it.
Three are block table bugs caught by C2 and C3.
N0 leaves old physical block indices in the worker's block table after
sliding-window eviction replaces the logical blocks.
N4 gives every parallel sample of one request the same disaggregation
transfer room,
so the samples' KV transfers land in one another's blocks.
N6 keeps a streaming session's recurrent state past the tokens the session
keeps,
after a speculative verify overshoots the limit or a finished turn runs once
more.
Two are entry access bugs caught by C5.
N3 quantizes a short request's never-written state slot into its int8
checkpoint.
N5 launches the freed row of a request that finished at the grammar
synchronization of the previous step.

These bugs range from silent unsafe transitions to visible failures.
N0 and N3 have no visible symptom on the current kernels,
which mask the stale or unwritten entries before consuming them,
whereas N2 leaves a request unanswered and N6 changes the next turn's output
distribution.
\sys reports all of them at the unsafe transition,
so a later kernel change cannot silently turn a masked one into output
corruption.
All eight are reported upstream with a fix pull request each,
listed as Fix PR in \autoref{t:eval-new-findings}.

\section{Discussion}
\label{s:discuss}

\PP{Generality Across Serving Systems}%
\sys's abstractions are not tied to one serving system,
and the evaluation on both vLLM and SGLang shows it.
\sys relies on three properties that any \PgdAttn-style serving system exposes:
a pool of fixed-size physical KV blocks carved out of a single GPU buffer,
a serving-managed block table that maps each request's logical blocks to
physical blocks,
and GPU kernels that consume this block table to compute KV addresses.
vLLM and SGLang realize these properties differently.
vLLM allocates 16-token blocks from a block pool and shares prefixes through
content hashes,
while SGLang allocates tokens or pages from a token pool and shares prefixes
through a radix tree,
and the two run different attention kernels.
The shadow block and the five checks (C1--C5) apply to both unchanged.
Porting \sys to a system consists of attaching C1 to its block manager,
C2 and C3 to its block table,
and C4 and C5 to its KV-access kernels.
The per-request permission field covers prefix sharing and speculative decoding
in both systems,
since shared and draft requests still allocate and access blocks through the
same block table.
Hybrid attention and SSM models add new KV-touching kernels,
which \sys instruments with the same helpers,
and the evaluated hybrid model runs under both systems this way.
Disaggregated serving extends block lifetime across nodes,
but each node keeps its own shadow blocks
and checks the transfer as a read or write (\autoref{ss:design-kv-transfer}),
so the abstraction itself does not change.

% \PP{Generality Beyond vLLM}%
% \sys is implemented in vLLM, but its abstractions are not tied to
% vLLM's codebase.
% %
% \sys relies on three properties common to paged KV-cache serving systems:
% GPU-side KV cache storage managed as reusable blocks or pages,
% serving-managed mappings from each request's token positions to those
% blocks or pages,
% and GPU kernels that consume those mappings to locate KV cache entries.
% %
% SGLang and TensorRT-LLM expose the same architectural pattern, although
% their concrete data structures differ:
% SGLang uses request-to-token and token-to-KV pools,
% while TensorRT-LLM uses a KV cache manager and paged KV-cache block
% offset metadata.
% %
% Thus, the shadow-block abstraction and checks C1--C5 carry over, but
% C2 and C3 must be attached to the target system's own mapping metadata,
% and C4 and C5 must be inlined into its KV-access kernels.
% %
% The per-request permission field also covers prefix caching because
% shared prefix blocks or pages can be treated as read-only for all
% requests that reuse them.
% %
% Speculative decoding can be handled by assigning permissions to the
% draft and verification KV state that the serving system exposes.
% %
% Hybrid attention and SSM models add new KV-touching kernels, which
% \sys requires to be instrumented with the same access-check helpers.
% %
% Disaggregated serving extends block lifetime across nodes and requires
% shadow-block metadata to be transferred together with the KV block
% contents, but the abstraction itself does not change.

\PP{Out-of-Scope Bugs}%
\sys targets KV cache allocation and access bugs,
not all model-quality or serving-correctness bugs.
Two classes of bugs fall outside this scope.
The first class is hash-collision bugs in prefix caching~\cite{wu2026cache},
which are wrong-sharing bugs caused by non-collision-free content hashes
rather than by violations of permission,
translation,
or access rules.
The second class is bugs in auxiliary metadata that is not part of KV cache
allocation or access, such as quantization scaling factors used by
reduced-precision kernels, or coordination messages exchanged between
distributed serving components.
These bugs can corrupt serving behavior,
but they do not violate the KV cache permission,
translation,
or access rules that \sys checks.

\PP{Use Cases}%
\sys is designed as a sanitizer rather than an always-on production guard,
because its GPU-side checks add a shadow block lookup and an initialization
check on the critical path of every KV cache access.
This cost positions \sys for the same usage pattern as ASan and Compute
Sanitizer,
both of which impose multi-fold slowdowns
(roughly 2\x for ASan and an order of magnitude for Compute Sanitizer
on GPU kernels)
yet remain widely used through four sanitizer-style use cases:
continuous integration,
pre-release validation,
post-incident debugging,
and sampled production.

The first use case is continuous integration, where \sys catches silent KV cache
corruptions or potential safety violations that current unit tests miss, such as
the eight new issues reported in \autoref{ss:eval-bug-detection}.
The second use case is pre-release validation of new kernels or serving
features,
where a developer runs \sys with every check enabled locally and verifies that
no check fires before merging.
The third use case is post-incident debugging of corrupted outputs,
where \sys reports a violation at the boundary that produced it
rather than at a downstream output many requests later.
The fourth use case is sampled production,
where \sys runs on a small fraction of machines or requests,
following the same strategy that ASan and Compute Sanitizer use at
fleet scale to amortize their cost;
a single violation report on a sampled request is sufficient to surface
a bug that would otherwise produce silent corruption across the fleet.

\section{Related work}
\label{s:relwk}

\PP{Bug Finding in LLM Serving Systems}%
GRIEF~\cite{grief} fuzzes vLLM and SGLang
with timed traces of concurrent requests,
but its oracles see a KV cache bug only through its symptoms.
GRIEF reports a bug when the engine crashes, hangs, or slows down,
when an output changes under log-probability replay,
or when KV block lifecycle events become inconsistent.
A KV cache safety violation, however, often shows none of these symptoms.
For example, N3 reads an uninitialized state slot that no request consumes,
and N7 frees blocks that an in-flight KV transfer still reads,
yet neither run crashes, hangs, or produces a wrong output.
Moreover, 23 of the 37 bugs in \autoref{t:motiv-bug-list}
are caught only at the GPU block table, its lookup, or the KV entry access (C3
to C5), which allocation and free events do not observe.
\sys instead checks each violation at the step that introduces it
and reports the check, block, and request involved.
A fuzzer such as GRIEF can therefore use \sys as its oracle
to find a broader range of KV cache safety bugs
and to triage each one from the involved check, block, and request.

\PP{Sanitizer for Custom Allocators}%
Custom allocators are common in high-performance CPU applications, and have
therefore been a target of memory safety research for decades.
ASan protects objects allocated through standard allocation APIs, but custom
allocators often allocate a large region and carve software-defined objects from
it.
As a result, object boundaries and object lifetimes inside an allocator can be
invisible to ASan.
CMASan addresses this gap by identifying custom allocator APIs,
instrumenting allocator operations,
and maintaining object metadata and quarantine state for objects created by
those allocators~\cite{hong2025cmasan}.
\sys follows the same high-level observation that the protected object may be
defined by software above the standard allocator.
However,
the KV cache setting requires different metadata and checks:
the protected objects are GPU-resident KV cache blocks,
and an access is safe only if the serving system grants the request the right
permission,
the GPU block table names the intended physical block,
and the target KV entry has been initialized.

\PP{Memory Safety for GPU Applications}%
% GPU memory safety tools detect memory errors in CUDA and GPU programs, but they
% operate at a lower abstraction than KV cache blocks.
%
% NVIDIA Compute Sanitizer detects invalid and uninitialized GPU memory accesses
% using CUDA-visible allocation information~\cite{nvidia-compute-sanitizer}.
% %
% This is effective for ordinary out-of-bounds accesses, but the tool sees a KV
% cache buffer as one large GPU allocation rather than seeing software-managed
% physical KV cache blocks.
% %
Prior GPU memory safety systems explore several points in detecting memory
errors in CUDA and GPU programs, but they operate at a lower abstraction than KV
cache blocks.
Canary-based tools such as GMOD~\cite{di2018gmod} and
clArmor~\cite{erb2018clarmor} detect some GPU buffer overflows with low
overhead, but can miss non-contiguous or non-linear corruptions.
Hardware-assisted systems such as GPUShield and LMI provide fine-grained bounds
metadata, but require hardware support or allocation constraints that limit
deployment on existing systems~\cite{lee2022gpushield,lee2025lmi}.
cuCatch instruments CUDA programs with pointer tags and bounds metadata, but
targets generic CUDA memory safety rather than serving-level KV cache
semantics~\cite{tarek2023cucatch}.
These systems protect GPU buffers from generic spatial or temporal memory
errors.
\sys is complementary and specialized: a KV cache bug can use an address that is
inside the allocated GPU buffer and still violate memory safety because the
kernel follows the wrong GPU block table entry, accesses a block without the
required permission, or reads an uninitialized KV entry.

\PP{KV Cache Security}%
Recent work on KV cache security shows that the KV cache opens a new attack
surface in LLM serving systems.
Cache Me, Catch You~\cite{wu2026cache} studies prefix, multimodal, and semantic
cache attacks, including collision-based prefix cache poisoning and cache-based
bypasses of system checks.
Other work studies side channels created by KV cache sharing in multi-tenant
serving.
I Know What You Asked~\cite{wu2025know} shows that an attacker can infer whether
its requests share KV cache with another user's request and can use that signal
to reconstruct prompts.
Both directions show that the KV cache is a security-critical component, while \sys
targets memory safety violations on the GPU-resident KV cache rather than
cache-key correctness, poisoning resistance, or side-channel leakage from cache
reuse.

\section{Conclusion}
\label{s:conclusion}

This paper presents \sys,
the first systematic memory-safety abstraction
for KV cache management in LLM serving systems.
\sys maintains a compact shadow block for each physical KV block
that records per-request permission and per-entry initialization,
and uses this metadata to perform five checks
along the \PgdAttn workflow,
catching each violation at the boundary that produced it
rather than as downstream output corruption many requests later.
%
% \BL{update number later}
% \JU{Updated.}
A prototype of \sys on vLLM and SGLang covers all 29 historical KV cache
bugs from the two systems' patch histories,
uncovers 8 previously unknown safety violations during end-to-end serving,
and reports no false positives.
\sys adds 3.0\% and 7.4\% geomean TPOT overhead
on Qwen 2.5 14B on vLLM and SGLang
and at most 0.11\% of the KV cache in memory overhead,
making \sys a practical sanitizer
for continuous integration,
pre-release validation,
and post-incident debugging of LLM serving systems.

% \input{ack}

% \appendix
% \input{appendix}

\bibliographystyle{plain}
% \footnotesize
% \setlength{\bibsep}{3pt}
\bibliography{p,sslab,conf}

\begin{thebibliography}{10}

\bibitem{mte}
{ARM Limited}.
\newblock {\em {Memory Tagging Extension}}, 2019.
\newblock
  \URL{https://developer.arm.com/-/media/Arm\%20Developer\%20Community/PDF/Arm_Memory_Tagging_Extension_Whitepaper.pdf}.

\bibitem{bai2024longbench}
Yushi Bai, Xin Lv, Jiajie Zhang, Hongchang Lyu, Jiankai Tang, Zhidian Huang,
  Zhengxiao Du, Xiao Liu, Aohan Zeng, Lei Hou, et~al.
\newblock {LongBench}: A bilingual, multitask benchmark for long context
  understanding.
\newblock In {\em Proceedings of the 62nd annual meeting of the association for
  computational linguistics (volume 1: Long papers)}, pages 3119--3137, 2024.

\bibitem{di2018gmod}
Bang Di, Jianhua Sun, Dong Li, Hao Chen, and Zhe Quan.
\newblock {GMOD}: A dynamic {GPU} memory overflow detector.
\newblock In {\em Proceedings of the 27th International Conference on Parallel
  Architectures and Compilation Techniques}, pages 1--13, 2018.

\bibitem{erb2018clarmor}
Christopher Erb and Joseph~L Greathouse.
\newblock {clArmor}: A dynamic buffer overflow detector for {OpenCL} kernels.
\newblock In {\em Proceedings of the International Workshop on OpenCL}, pages
  1--2, 2018.

\bibitem{gao2024cost}
Bin Gao, Zhuomin He, Puru Sharma, Qingxuan Kang, Djordje Jevdjic, Junbo Deng,
  Xingkun Yang, Zhou Yu, and Pengfei Zuo.
\newblock {Cost-Efficient} large language model serving for multi-turn
  conversations with {CachedAttention}.
\newblock In {\em 2024 USENIX annual technical conference (USENIX ATC 24)},
  pages 111--126, 2024.

\bibitem{grattafiori2024llama}
Aaron Grattafiori, Abhimanyu Dubey, Abhinav Jauhri, Abhinav Pandey, Abhishek
  Kadian, Ahmad Al-Dahle, Aiesha Letman, Akhil Mathur, Alan Schelten, Alex
  Vaughan, et~al.
\newblock The llama 3 herd of models.
\newblock {\em arXiv preprint arXiv:2407.21783}, 2024.

\bibitem{hendrycks2021measuring}
Dan Hendrycks, Collin Burns, Steven Basart, Andy Zou, Mantas Mazeika, Dawn
  Song, and Jacob Steinhardt.
\newblock Measuring massive multitask language understanding.
\newblock In {\em International Conference on Learning Representations}, 2021.

\bibitem{hong2025cmasan}
Junwha Hong, Wonil Jang, Mijung Kim, Lei Yu, Yonghwi Kwon, and Yuseok Jeon.
\newblock Cmasan: Custom memory allocator-aware address sanitizer.
\newblock In {\em 2025 IEEE Symposium on Security and Privacy (SP)}, pages
  740--757. IEEE, 2025.

\bibitem{vllm}
Woosuk Kwon, Zhuohan Li, Siyuan Zhuang, Ying Sheng, Lianmin Zheng, Cody~Hao Yu,
  Joseph~E. Gonzalez, Hao Zhang, and Ion Stoica.
\newblock Efficient memory management for large language model serving with
  {PagedAttention}.
\newblock In {\em Proceedings of the 29th symposium on operating systems
  principles}, pages 611--626, 2023.

\bibitem{lee2025lmi}
Jaewon Lee, Euijun Chung, Saurabh Singh, Seonjin Na, Yonghae Kim, Jaekyu Lee,
  and Hyesoon Kim.
\newblock {Let-Me-In}: ({Still}) employing in-pointer bounds metadata for
  fine-grained {GPU} memory safety.
\newblock In {\em 2025 IEEE International Symposium on High Performance
  Computer Architecture (HPCA)}, pages 1648--1661. IEEE, 2025.

\bibitem{lee2022gpushield}
Jaewon Lee, Yonghae Kim, Jiashen Cao, Euna Kim, Jaekyu Lee, and Hyesoon Kim.
\newblock Securing {GPU} via region-based bounds checking.
\newblock In {\em Proceedings of the 49th Annual International Symposium on
  Computer Architecture}, pages 27--41, 2022.

\bibitem{lee2024infinigen}
Wonbeom Lee, Jungi Lee, Junghwan Seo, and Jaewoong Sim.
\newblock {InfiniGen}: Efficient generative inference of large language models
  with dynamic {KV} cache management.
\newblock In {\em 18th USENIX Symposium on Operating Systems Design and
  Implementation (OSDI 24)}, pages 155--172, 2024.

\bibitem{softbound}
Santosh Nagarakatte, Jianzhou Zhao, Milo M.~K. Martin, and Steve Zdancewic.
\newblock {SoftBound}: Highly compatible and complete spatial memory safety for
  {C}.
\newblock In {\em Proceedings of the 30th ACM SIGPLAN Conference on Programming
  Language Design and Implementation (PLDI)}, pages 245--258, 2009.

\bibitem{cets}
Santosh Nagarakatte, Jianzhou Zhao, Milo M.~K. Martin, and Steve Zdancewic.
\newblock {CETS}: Compiler-enforced temporal safety for {C}.
\newblock In {\em Proceedings of the 2010 International Symposium on Memory
  Management (ISMM)}, pages 31--40, 2010.

\bibitem{nvidia-tensorrt-llm}
{NVIDIA Corporation}.
\newblock {\em {TensorRT-LLM}}, 2025.
\newblock \URL{https://github.com/NVIDIA/TensorRT-LLM}.

\bibitem{nvidia-compute-sanitizer}
{NVIDIA Corporation}.
\newblock {\em {Compute Sanitizer Documentation}}, 2026.
\newblock
  \URL{https://docs.nvidia.com/compute-sanitizer/ComputeSanitizer/index.html}.

\bibitem{nvtx}
{NVIDIA Corporation}.
\newblock {\em {NVTX API for Compute Sanitizer Reference Manual}}, 2026.
\newblock
  \URL{https://docs.nvidia.com/compute-sanitizer/SanitizerNvtxGuide/index.html}.

\bibitem{kvcache}
Reiner Pope, Sholto Douglas, Aakanksha Chowdhery, Jacob Devlin, James Bradbury,
  Anselm Levskaya, Jonathan Heek, Kefan Xiao, Shivani Agrawal, and Jeff Dean.
\newblock Efficiently scaling transformer inference.
\newblock {\em Proceedings of machine learning and systems}, 5:606--624, 2023.

\bibitem{qin2026mooncake}
Ruoyu Qin, Zheming Li, Weiran He, Jialei Cui, Heyi Tang, Feng Ren, Teng Ma,
  Shangming Cai, Yineng Zhang, Mingxing Zhang, et~al.
\newblock Mooncake: A {KVCache}-centric disaggregated architecture for {LLM}
  serving.
\newblock {\em ACM Transactions on Storage}, 22(4):1--38, 2026.

\bibitem{sharegpt52k}
{RyokoAI}.
\newblock {\em {ShareGPT52K Dataset}}, 2023.
\newblock \URL{https://huggingface.co/datasets/RyokoAI/ShareGPT52K}.

\bibitem{asan}
Konstantin Serebryany, Derek Bruening, Alexander Potapenko, and Dmitry Vyukov.
\newblock {AddressSanitizer}: A fast address sanity checker.
\newblock In {\em 2012 USENIX Annual Technical Conference (USENIX ATC 12)},
  pages 309--318, 2012.

\bibitem{shah2024flashattention}
Jay Shah, Ganesh Bikshandi, Ying Zhang, Vijay Thakkar, Pradeep Ramani, and Tri
  Dao.
\newblock Flashattention-3: Fast and accurate attention with asynchrony and
  low-precision.
\newblock {\em Advances in Neural Information Processing Systems},
  37:68658--68685, 2024.

\bibitem{sheng2023flexgen}
Ying Sheng, Lianmin Zheng, Binhang Yuan, Zhuohan Li, Max Ryabinin, Beidi Chen,
  Percy Liang, Christopher R{\'e}, Ion Stoica, and Ce~Zhang.
\newblock {FlexGen}: High-throughput generative inference of large language
  models with a single {GPU}.
\newblock In {\em International Conference on Machine Learning}, pages
  31094--31116. PMLR, 2023.

\bibitem{tarek2023cucatch}
Mohamed Tarek Ibn~Ziad, Sana Damani, Aamer Jaleel, Stephen~W Keckler, and Mark
  Stephenson.
\newblock {cuCatch}: A debugging tool for efficiently catching memory safety
  violations in {CUDA} applications.
\newblock {\em Proceedings of the ACM on Programming Languages},
  7(PLDI):124--147, 2023.

\bibitem{cheri}
Robert N.~M. Watson, Jonathan Woodruff, Peter~G. Neumann, Simon~W. Moore,
  Jonathan Anderson, David Chisnall, Nirav Dave, Brooks Davis, Khilan Gudka,
  Ben Laurie, et~al.
\newblock {CHERI}: A hybrid capability-system architecture for scalable
  software compartmentalization.
\newblock In {\em 2015 IEEE Symposium on Security and Privacy (SP)}, pages
  20--37. IEEE, 2015.

\bibitem{wu2025know}
Guanlong Wu, Zheng Zhang, Yao Zhang, Weili Wang, Jianyu Niu, Ye~Wu, and Yinqian
  Zhang.
\newblock I know what you asked: Prompt leakage via {KV}-cache sharing in
  multi-tenant {LLM} serving.
\newblock In {\em NDSS}, 2025.

\bibitem{wu2026cache}
XiangFan Wu, Lingyun Ying, Guoqiang Chen, Yacong Gu, and Haipeng Qu.
\newblock Cache me, catch you: Cache related security threats in {LLM} serving
  frameworks.
\newblock NDSS, 2026.

\bibitem{yang2024qwen25}
An~Yang, Baosong Yang, Beichen Zhang, Binyuan Hui, Bo~Zheng, Bowen Yu,
  Chengyuan Li, Dayiheng Liu, Fei Huang, Haoran Wei, et~al.
\newblock {Qwen2.5} technical report.
\newblock {\em arXiv preprint arXiv:2412.15115}, 2024.

\bibitem{zhang2025jenga}
Chen Zhang, Kuntai Du, Shu Liu, Woosuk Kwon, Xiangxi Mo, Yufeng Wang, Xiaoxuan
  Liu, Kaichao You, Zhuohan Li, Mingsheng Long, et~al.
\newblock Jenga: Effective memory management for serving {LLM} with
  heterogeneity.
\newblock In {\em Proceedings of the ACM SIGOPS 31st Symposium on Operating
  Systems Principles}, pages 446--461, 2025.

\bibitem{grief}
Yunze Zhao, Yibo Zhao, Yuchen Zhang, Zaoxing Liu, and Michelle~L Mazurek.
\newblock Continuous discovery of vulnerabilities in {LLM} serving systems with
  fuzzing.
\newblock {\em arXiv preprint arXiv:2605.11202}, 2026.

\bibitem{sglang}
Lianmin Zheng, Liangsheng Yin, Zhiqiang Xie, Chuyue Sun, Jeff Huang, Cody~Hao
  Yu, Shiyi Cao, Christos Kozyrakis, Ion Stoica, Joseph~E Gonzalez, et~al.
\newblock {SGLang}: Efficient execution of structured language model programs.
\newblock {\em Advances in neural information processing systems},
  37:62557--62583, 2024.

\bibitem{zhong2024distserve}
Yinmin Zhong, Shengyu Liu, Junda Chen, Jianbo Hu, Yibo Zhu, Xuanzhe Liu, Xin
  Jin, and Hao Zhang.
\newblock {DistServe}: Disaggregating prefill and decoding for
  goodput-optimized large language model serving.
\newblock In {\em 18th USENIX Symposium on Operating Systems Design and
  Implementation (OSDI 24)}, pages 193--210, 2024.

\end{thebibliography}

\end{document}